\documentclass[prd, twocolumn, amsmath,amssymb,aps, reprint, preprintnumbers, nofootinbib, superscriptaddress]
{revtex4-2}
\usepackage{xcolor}
\usepackage{amsmath}
\usepackage{physics}
\usepackage{amssymb}
\usepackage{graphicx}
\usepackage{svg}
\usepackage[toc,page]{appendix}
\usepackage{url}
\usepackage{ragged2e}
\pdfoutput=1
\usepackage{amsfonts}
\usepackage{mathrsfs}
\usepackage{graphicx}
\usepackage{hyperref}
\hypersetup{colorlinks=true,allcolors=blue}
\usepackage{makecell}

\def\mC{{\mathcal{C}}}
\def\mO{{\mathcal{O}}}
\def\tempC{{C}}
\def\tempO{{O}}
\def\mc{{\textbf{c}}}
\def\mo{{\textbf{o}}}

\def \thl {{\theta_l}}
\def \thV {{\theta_{V}}}

\begin{document}

\begin{abstract}
    The $SU(2)_L\times U(1)_Y$ invariance of the Standard Model Effective Field Theory (SMEFT) imposes relations among different low-energy effective field theory Wilson coefficients (WCs), any deviations from which would signal the presence of physics beyond SMEFT. In this work, we investigate two such relations ($\mC_{S} = -\,\mC_{P}$, $\mC_S^\prime = \mC_{P}^\prime$) among the scalar and the pseudoscalar new-physics WCs that can contribute to $b\to s \,\tau \tau $ processes.
    We show that, even when new physics violating these relations would not be measurable in the branching ratios of $B_s \to \tau^+ \tau^-$ and  $B\to K^{(*)}\,\tau^+ \tau^-$ at HL-LHC and FCC-ee, it can still manifest itself in the angular distribution of $B\to K^{*0}\,\tau^+ \tau^-$.
    We identify the combinations of angular observables in this decay channel that are sensitive to scenarios beyond SMEFT. We find that the two observables $S_6^c$ and $A_7$, and their combination, have the potential to identify physics beyond SMEFT. 
\end{abstract}

\author{Siddhartha Karmakar}
\email{siddhartha@theory.tifr.res.in (ORCID: 0009-0003-0609-9689)}
\affiliation{Tata Institute of Fundamental Research, Homi Bhabha Road, Colaba, Mumbai 400005, India}
\author{Amol Dighe} 
\email{amol@theory.tifr.res.in (ORCID: 0000-0001-6639-0951)}
\affiliation{Tata Institute of Fundamental Research, Homi Bhabha Road, Colaba, Mumbai 400005, India}
\preprint{TIFR/TH/24-16}
\title{Exploring observable effects of scalar operators beyond SMEFT\\ in the angular distribution of $B\to K^{*0} \tau^+ \tau^- $}
\maketitle

\section{Introduction}

Physics beyond the Standard Model (BSM) is often parameterized model-independently in terms of the Standard Model Effective Field Theory (SMEFT)~\cite{Grzadkowski:2010es, Isidori:2023pyp}. The effective Lagrangian in SMEFT is written as
\begin{align}
    \mathcal{L}&= \mathcal{L}_{SM}+\frac{1}{\Lambda}{\tempC}^{(5)}O^{(5)}+\frac{1}{\Lambda^2}\sum_{i}C_i^{(6)}O_i^{(6)}+\mathcal{O}\left(\frac{1}{\Lambda^3}\right).\label{SMEFTexpansion}
\end{align}
Here ${\tempO}_i^{(d)}$ are the effective operators with mass dimension $d > 4$ which are comprised of the Standard Model (SM) fields, and the Wilson coefficients (WCs) ${\tempC}_i^{(d)}$ encode the corresponding new physics (NP) effects. All the operators in SMEFT are invariant under the SM gauge symmetry $SU(3)_C\times SU(2)_L\times U(1)_Y$.
 
In flavor physics, the relevant energy scale is near the mass of the $B$ meson. At this scale, the heavier particles, i.e. $Z$, $W^{\pm}$ gauge bosons, the Higgs boson ($h$) and the top quark, are integrated out. The resulting effective field theory is the Low-energy Effective Field Theory (LEFT)~\cite{Buchalla:1995vs, Jenkins:2017jig}. As LEFT is relevant below the electroweak (EW) scale, all the operators in LEFT need to be invariant only under $SU(3)_C\times U(1)_{em}$. When LEFT operators are matched to SMEFT operators, the $SU(2)_L\times U(1)_Y$ invariance of SMEFT imposes several relations among the WCs in LEFT. Such relations and their phenomenological implications have recently been studied in \cite{Alonso:2014csa, Cata:2015lta, Azatov:2018knx, Fuentes-Martin:2020lea, Bause:2020auq, Bause:2020xzj, Bissmann:2020mfi, Bause:2021cna, Bause:2021ihn, Bruggisser:2021duo, Bause:2022rrs, Sun:2023cuf, Grunwald:2023nli, Greljo:2023bab, Fajfer:2012vx, Bause:2023mfe, Bhattacharya:2023beo, Chen:2024jlj, Fernandez-Martinez:2024bxg, Karmakar:2024gla}. 
 
However, SMEFT is not the only effective field theory (EFT) above the EW scale. There are more general EFTs such as the Higgs Effective Field Theory (HEFT)~\cite{Alonso:2012px, Buchalla:2013rka, Pich:2016lew, Cohen:2020xca, Burgess:2021ylu} where the EW symmetry $SU(2)_L\times U(1)_Y$ is non-linearly realized. If HEFT is the effective theory above the EW scale, then the relations predicted by SMEFT may no longer be valid. Although SMEFT is a more commonly used EFT in literature, there is no experimental evidence yet to prefer it over HEFT. Searches for beyond-SMEFT physics in ATLAS and CMS focus on the precise measurements of Higgs couplings with fermions and gauge bosons~\cite{DiMicco:2019ngk}. In this work, our goal is to study the possibility of identifying the effects of physics beyond SMEFT in the $b\to s \tau\tau$ channel.  

The motivation for focusing on the $b\to \tau$ sector emerges from several recently observed flavor anomalies in 
$B$ meson decays that indicate the possibility of BSM physics. For example, the measurements of $R(D^{(*)})$~\cite{BaBar:2012obs,BaBar:2013mob,Belle:2015qfa,LHCb:2015gmp} and  $R(J/\psi)$~\cite{LHCb:2017vlu} involve the charged-current transition $b\rightarrow c \tau \nu_\tau$, while the observables $\mathcal{B}(B^+\rightarrow K^+\mu^+\mu^-)$~\cite{LHCb:2014cxe}, $\mathcal{B}(B^+\rightarrow K^+e^+e^-)$~\cite{LHCb:2022zom}, ${\cal B}(B\to K \nu \bar \nu)$~\cite{Belle-II:2023esi} and $P^\prime_5$~\cite{LHCb:2013ghj, Descotes-Genon:2012isb,Descotes-Genon:2013wba} involve the neutral-current transitions $b\rightarrow s \ell \ell$.
Although these anomalies can be addressed within the framework of SMEFT, the possibility that they originate from physics beyond SMEFT remains open. 
In our earlier study~\cite{Karmakar:2023rdt}, we have explored the possibility of identifying the beyond-SMEFT effects in the charged-current transition $b\rightarrow c \tau \nu_\tau$, using angular observables in $\Lambda_b\to \Lambda_c(\to \Lambda\pi)\tau\bar\nu_\tau$ and constraints from $R(D^{(*)})$, $R(J/\psi)$, ${\cal B}(B_c\to \tau^+\tau^-)$.

In this work, we focus on the processes mediated by $b\to s \tau\tau$. 
In recent literature~\cite{Alonso:2015sja, Crivellin:2017zlb, Calibbi:2017qbu, Capdevila:2017iqn, Bause:2021cna, Bause:2023mfe, Allwicher:2023xba, Karmakar:2024gla}, it has been pointed out that a common explanation of observed excess in ${\cal B}(B^+\to K^+\nu \bar\nu)$ and the deviations in the lepton flavor universality (LFU) ratios $R(D^{(*)})$ would predict excess branching fractions for the modes $B_s\to \tau^+\tau^-$, $B\to K^{(*)}\tau^+\tau^-$, etc. 
It should be noted that, in addition to the branching fractions, the NP will also affect the angular distributions in the $B\to K^{(*)}\tau^+\tau^-$ modes.
In fact, even in the cases where the NP effects are not evident in the branching ratios or are obscured by hadronic uncertainties, they may still show up in the angular distributions. In our analysis, we study the possibility of identifying the NP effects -- specifically the effects beyond SMEFT -- in the angular observables of $B\to K^{(*)}\tau^+\tau^-$, even when they are not measurable in the branching ratios.  

Two of the SMEFT predicted relations among the WCs of scalar and pseudoscalar LEFT operators are
\begin{align}
    \mC_S &= -\mC_P \quad {\rm and} \quad \mC_S^\prime = \mC_P^\prime~.
\end{align}
We study the possible violation of these relations and the prospects of identifying the physics beyond SMEFT through future measurements of the processes $B_s\to \tau^+\tau^-$ and $B\to K^{(*)}\tau^+\tau^-$, which may get contributions from the semileptonic scalar and pseudoscalar operators in the $b\to s \tau\tau$ channel. 

Currently, the branching ratios of these modes are not well measured. The present experimental bounds are ${\cal B}(B_s\to \tau^+\tau^-)<6.8\times10^{-3}$~\cite{LHCb:2017myy}, ${\cal B}(B^+\to K^+ \tau^+\tau^-)<2.25\times10^{-3}$~\cite{BaBar:2016wgb} and ${\cal B}(B^0\to K^{*0} \tau^+\tau^-)<3.1\times10^{-3}$~\cite{Belle:2021ecr}. These bounds are much weaker as compared to the SM expectations, which are ${\cal O}(10^{-7})$ for all these modes~\cite{Straub:2018kue}. However, these bounds will improve by one or two orders of magnitude in Belle-II~\cite{Belle-II:2018jsg} and HL-LHC~\cite{LHCb:2018roe}. Moreover, FCC-ee is expected to measure ${\cal B}(B \to K^{(*)} \tau^+\tau^-)$ at the SM level with the estimated yield for  $B \to K^{*} \tau^+\tau^-$ of ${\cal O}(1000)$ in the first two years of running (phase - 1)~\cite{FCC:2018byv, Apollonio:2019zjt}. This will allow us to carry out angular analyses for these modes.

In this work, we envisage the scenario where the NP effects may not be apparent directly in the branching ratios of $B_s\to \tau^+\tau^-$ and $B\to K^{(*)}\tau^+\tau^-$ at the level of precision possible at HL-LHC and FCC-ee. Taking these branching ratios to be consistent with the SM, we calculate constraints for the WCs in LEFT contributing to $b\to s \tau\tau$ processes based on the projected measurements of these branching ratios. Given these constraints, we explore if the angular observables in $B \to K^{*0} \tau^+\tau^-$ are sensitive to any deviations from the SMEFT-predicted relations. We look for combinations of these angular observables that will be able to identify physics beyond SMEFT. Note that angular distributions of $B\to K^{*0}\tau^+\tau^-$ have been studied earlier in \cite{Aliev:2000ae,Choudhury:2003mi, SinghChundawat:2022ldm}; however, their potential for identifying beyond-SMEFT effects has not yet been explored.

In Sec.~\ref{sec: eft}, we list the effective operators in SMEFT and in LEFT that contribute to the $b\to s\tau\tau$ processes and discuss the predictions of SMEFT for the LEFT WCs. In Sec.~\ref{sec: bounds}, we present the projected constraints on the WCs in LEFT with and without the underlying SMEFT assumptions, taking the NP WCs to be real. In Sec.~\ref{sec:results}, we discuss the angular observables in $B\to K^{*0}\tau^+\tau^-$ and try to identify suitable observables, and their combinations, for distinguishing the scenarios within and beyond SMEFT. In Sec.~\ref{sec:complex}, we extend this exercise to the case of complex WCs, pointing out an additional CP-odd observable sensitive to physics beyond SMEFT. Finally, we summarize our results in Sec.~\ref{sec:conclusion}.

\section{SMEFT-predicted relations for Wilson coefficients in LEFT}\label{sec: eft}

In this section, we list the dimension-6 semileptonic operators in LEFT contributing to the $b\to s\tau\tau$ transition at the tree level. Based on the matching of these operators to SMEFT, we present the resulting relations among the WCs in LEFT. 

The leading order effective Lagrangian in LEFT for $b\rightarrow s \tau^+\tau^-$ process, considering only semileptonic operators, is~\cite{Das:2018iap}
\begin{equation}\label{Leff}
    \mathcal{L}^{\rm eff} = {\cal L}_{\rm SM} +  \frac{4G_F}{\sqrt{2}}V_{tb}V_{ts}^\ast\frac{\alpha_e}{4\pi}  \sum_i \mC_i \mathcal{O}_i \, ,
\end{equation}
where the operators ${\mO}_i$ include the following 
\begin{align}
    &\mathcal{O}^{(\prime)}_9 = \big[\bar{s}\gamma^\mu P_{L(R)}b \big]\big[\tau\gamma_\mu\tau \big]\, ,\quad \mathcal{O}_P^{(\prime)} = \big[\bar{s}P_R(L)b \big]\big[\tau\gamma_5\tau \big]\, ,\nonumber\\
    &\mathcal{O}_S^{(\prime)} = \big[\bar{s}P_R(L)b \big]\big[\tau\tau \big]\, , \quad \mathcal{O}^{(\prime)}_{10} = \big[\bar{s}\gamma^\mu P_{L(R)}b \big]\big[\tau\gamma_\mu\gamma_5\tau \big]\, ,\nonumber\\
    &\mathcal{O}_T = \big[\bar{s} \sigma^{\mu\nu} b \big]\big[\tau \sigma_{\mu\nu}\tau \big]\, ,\quad \mathcal{O}_{T5} = \big[\bar{s}\sigma^{\mu\nu} b \big]\big[\tau\sigma_{\mu\nu}\gamma_5\tau \big]\, .
\end{align}

Semileptonic operators in SMEFT contributing to $b\to s \tau\tau$ channel are
\begin{align}
    {\tempO}_{lq}^{(1)} &= (\bar q \gamma^\mu q)(\bar \ell \gamma_\mu \ell)~, \quad     \tempO_{lq}^{(3)} = (\bar q \gamma^\mu \tau_Iq)(\bar \ell \gamma_\mu \tau^I \ell)~, \\
    {\tempO}_{qe} &= (\bar q \gamma^\mu q)(\bar e \gamma_\mu e)~, \quad     \tempO_{ld} = (\bar d \gamma^\mu d)(\bar \ell \gamma_\mu \ell)~, \\
    {\tempO}_{ledq} &= (\bar \ell \,e )(\bar d \, q)~,
\end{align}
where $q$ and $\ell$ are left-handed quark and lepton doublets, whereas $d$ and $e$ are right-handed down-type quark and right-handed electron, respectively. 
The matching equations among the WCs of scalar and tensor operators in SMEFT and LEFT are~\cite{Cata:2015lta, Aebischer:2015fzz}
\begin{align}
    {\mC}_{S}^{\alpha\beta i j} &= \frac{v^2}{4\Lambda^2} {\tempC}_{ledq}^{\beta\alpha j i \, *}~,\quad  {\mC}_{P}^{\alpha\beta i j} =-\frac{v^2}{4\Lambda^2} {\tempC}_{ledq}^{\beta\alpha j i \, *}~,\\
    {\mC}_{S}^{\prime\,\alpha\beta i j} &= \frac{v^2}{4\Lambda^2} {\tempC}_{ledq}^{\alpha\beta i j}~,\quad  {\mC}_{P}^{\prime\,\alpha\beta i j} =\frac{v^2}{4\Lambda^2} {\tempC}_{ledq}^{\alpha\beta  i j\, *}~,\\
    {\mC}_T^{\alpha \beta i j} &= {\mC}^{\alpha\beta i j}_{T5}= 0~.
\end{align}
These matching equations imply the following relations~\cite{Cata:2015lta, Aebischer:2015fzz, Karmakar:2024gla}:
\begin{align}
    \mC_{S}&=-\mC_P\,,\quad \mC_{S}^{\prime}=\mC_P^{\prime}\,,\quad {\mC}_T = {\mC}_{T5}= 0~.\label{SPrelation}
\end{align}
These relations result from the conservation of $U(1)_Y$ hypercharge in the corresponding SMEFT operators. When considering operators up to dimension 6, these relations hold true irrespective of which SMEFT operators are present at the UV scale. However, in a more general EFT above the EW scale, such as HEFT, the EW gauge symmetry $SU(2)_L\times U(1)_Y$ is realized non-linearly, and therefore, the $U(1)_Y$ hypercharge need not be conserved in each HEFT operator separately. In this scenario, the resulting LEFT operators may not follow the SMEFT-predicted relations. The matching of LEFT scalar operators to such a non-linear EFT is detailed in \cite{Cata:2015lta}:
\begin{align}
    \mC_{S,P} &= \mathcal{N}(\pm \mc_S+ \mc_{Y1})~,\quad 	\mC_{S,P}^\prime = \mathcal{N}( \mc_S^\prime\pm \mc_{Y1}^\prime)~,\label{non-linear-map1}
\end{align}
where ${\cal N}$ is a normalization constant. Here $\mc_S^{(\prime)}$ and $\mc_{Y1}^{(\prime)}$ are the WCs of the corresponding  operators $\mo_S^{(\prime)}$ and $\mo_P^{(\prime)}$ in the non-linear EFT basis  in unitary gauge:
\begin{align}
    [{\mo}_S]^{\alpha\beta ij} &= -4\,(\bar d_L^i\gamma_\mu e_L^\alpha)(\bar e_R^\beta \gamma^\mu d_R^j) ~,\\
    [{\mo}_S^\prime]^{\alpha\beta ij} &= -4\,(\bar e_L^\beta \gamma_\mu d_L^j)(\bar d_R^i \gamma^\mu e_R^\alpha) ~,\\
    [{\mo}_{Y1}]^{\alpha\beta ij} &= (\bar e_L^\alpha\,e_R^\beta)(\bar d_L^i \, d_R^j)~,\\
    [{\mo}_{Y1}^\prime]^{\alpha\beta ij} &= (\bar e_R^\alpha\,e_L^\beta)(\bar d_R^i \, d_L^j)~.
\end{align}
Here $(\alpha, \beta)$ and $(i, j)$ correspond to the generation indices for leptons and quarks, respectively. The implication of eq.\,(\ref{non-linear-map1}) is that the relations in eq.\,(\ref{SPrelation}) are no longer maintained. The deviations from the correlations arise due to the WCs $ \mc_{Y1}$ and $ \mc_{Y1}^\prime$ and can be parameterized as
\begin{align}
    \Delta \mC &\equiv \mC_S +\mC_P = 2\mathcal{N}\, \mc_{Y1},\label{deltac}\\
    \Delta \mC^\prime &\equiv \mC_S^\prime -\mC_P^\prime = 2\mathcal{N}\, \mc_{Y1}^\prime~.\label{deltacp}
\end{align}	
Non-zero values of $\Delta \mC $ and $\Delta \mC^\prime$ indicate effects beyond SMEFT. For the tensor operators, the WCs ${\mC}_T$ and ${\mC}_{T5}$ are zero in SMEFT, so any nonzero value of these WCs will suggest physics beyond SMEFT. In our numerical analysis, we focus exclusively on the NP vector and scalar operators.

\begin{table*}[t]
    \centering
    \renewcommand{\arraystretch}{1.2}
    \begin{tabular}{|c|c|c|c|}
        \hline
        Observables & SM~\cite{Straub:2018kue} & Current bounds & HL-LHC + FCC-ee~\cite{FCC:2018byv}\\
        \hline
        ${\cal B}(B_s\to \tau^+\tau^-)$ & $(7.73\pm 0.49)\times 10^{-7}$ & $<6.8\times10^{-3}$~\cite{LHCb:2017myy} & $<10^{-5}$\\
        ${\cal B}(B^{+}\to K^{+}\tau^+\tau^-)$ & $(1.66\pm 0.17)\times 10^{-7}$ & $<2.25\times10^{-3}$~\cite{BaBar:2016wgb}& ${\rm SM}\pm 10\%$ \\
        ${\cal B}(B^{+}\to K^{*+}\tau^+\tau^-)$ & $(1.55\pm 0.20)\times 10^{-7}$ & $<3.1\times10^{-3}$~\cite{Belle:2021ecr} &${\rm SM}\pm 10\%$ \\
        ${\cal B}(B^{+}\to K^{*0}\tau^+\tau^-)$ & $(1.43\pm 0.18)\times 10^{-7}$ & $<3.1\times10^{-3}$~\cite{Belle:2021ecr}& ${\rm SM}\pm 10\%$\\
        \hline
    \end{tabular}
    \caption{\justifying SM values (column 2), current bounds (column 3) and expected precisions after HL-LHC and the phase-I of FCC-ee (column 4) for the branching ratios of $B_s\to \tau^+\tau^-$ and $B\to K^{(*)}\tau^+\tau^-$.}\label{tab: current_status}
\end{table*}

\section{Constraints on the LEFT WC's} \label{sec: bounds}

\begin{figure}[b]
    \centering
    \includegraphics[width=0.48\textwidth]{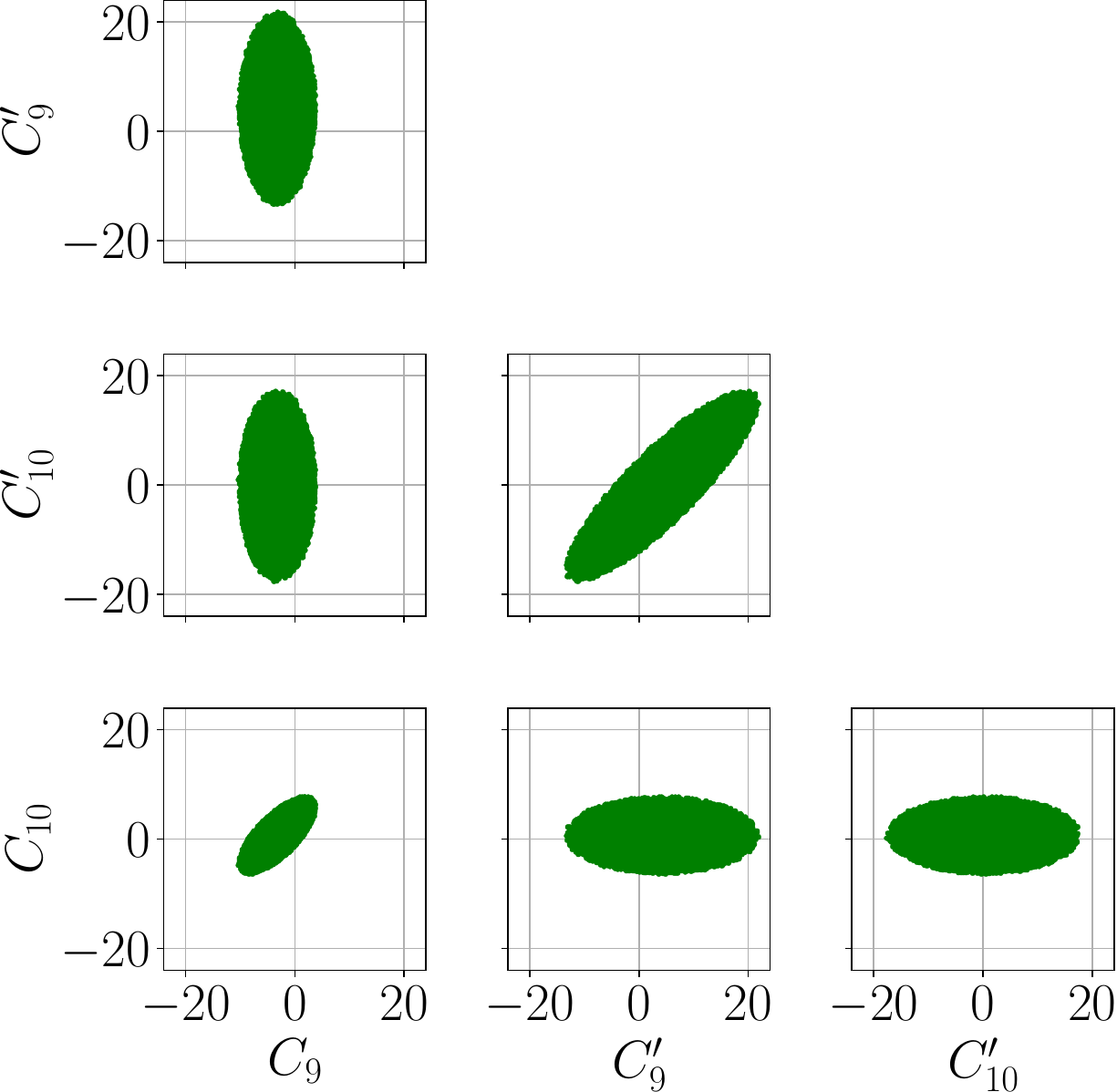}
    \caption{\justifying Projected bounds on $C_9$, $C_{10}$, $C_{9}^\prime$ and $C_{10}^\prime$ from the expected observation of ${\cal B}(B\to K^{*}\tau\tau)$ at FCC-ee to be within 10\% of its SM value. While considering any combination of two parameters, the remaining parameters are varied over all their possible values.}\label{fig:VAregion}
\end{figure}

The NP contributions to $b\to s\tau\tau$ transition may arise from the vector and scalar Wilson coefficients $\mC_{9}^{(\prime)}$, $\mC_{10}^{(\prime)}$, $C_{S}^{(\prime)}$ and $C_{P}^{(\prime)}$. To put constraints on these WCs, we consider the branching ratios ${\cal B}(B_s\to \tau^+\tau^-)$ and ${\cal B}(B\to K^{(*)}\tau^+\tau^-)$. The SM predictions, current upper bounds and projected future bounds for these modes are presented in Table~\ref{tab: current_status}. We envisage the scenario where the NP effects may not be apparent directly in the branching ratios of $B\to K^{(*)}\tau^+\tau^-$; however, they may affect the angular distribution of these decays. Based on these projected measurements, we calculate constraints on the WCs, both in the SMEFT and beyond-SMEFT scenarios. 

For the WCs associated with the vector operators, the best constraints come from  ${\cal B}(B\to K^{*}\tau^+\tau^-)$. While constraining these WCs corresponding to the vector operators, we consider the ``VA" scenario where we take all the scalar WCs to be zero. Since ${\cal O}(1000)$ such events are expected in the phase-I of FCC-ee~\cite{FCC:2018byv}, we project the future precision on ${\cal B}(B\to K^{(*)}\tau^+\tau^-)$ conservatively to be within 10\% of its SM value. 
That is, we estimate constraints on the vector WCs with the condition 
\begin{align}
    |{\cal B}_{\rm NP} - {\cal B}_{\rm SM}| < 0.1 \times {\cal B}_{\rm SM}~,
\end{align}
where ${\cal B}$ stands for ${\cal B}(B\to K^{(*)}\tau^+\tau^-)$. We use the python package {\tt flavio}~\cite{Straub:2018kue} to calculate the theoretical prediction for ${\cal B}(B\to K^{(*)}\tau^+\tau^-)$. The resulting constraints on the WCs $C_9$, $C_{10}$, $C_9^\prime$ and $C_{10}^{\prime}$ are shown in Fig.~\ref{fig:VAregion}.

For the scalar operators, we define
\begin{align}
    \mC_{S+} &\equiv \mC_S + \mC_S^\prime~, \quad \mC_{P+} \equiv \mC_P + \mC_P^\prime~,\\
    \mC_{S-} &\equiv \mC_S - \mC_S^\prime~, \quad \mC_{P-} \equiv \mC_P - \mC_P^\prime~.
\end{align}
In terms of the above quantities, the SMEFT-predicted relations for the WCs of the scalar operators become 
\begin{align}
    {\mC}_{S+} = - {\mC}_{P-} \quad {\rm and} \quad {\mC}_{P+} = - {\mC}_{S-}~.\label{SMEFTpedplusminus}
\end{align}
When these relations are obeyed, we call the scenario ``SP". 
Violations of the SMEFT-predicted relations can be parameterized in terms of the quantities $\Delta\mC_{1}$ and $\Delta\mC_{2}$, where
\begin{align}
    \Delta\mC_1 &\equiv {\mC}_{S+} + {\mC}_{P-}~, \quad \Delta\mC_2 \equiv {\mC}_{P+} + {\mC}_{S-}~.
\end{align} 
Note that $\Delta\mC_{1}$ and $\Delta\mC_{2}$ can be related to the quantities $\Delta\mC$ and $\Delta\mC^\prime$ defined in eqs.\,(\ref{deltac}) and (\ref{deltacp}) as
\begin{align}
    \Delta\mC_1 &\equiv \Delta\mC + \Delta\mC^\prime~, \quad \Delta\mC_2 \equiv \Delta\mC - \Delta\mC^\prime~.
\end{align}
When $\Delta\mC_{1}\neq 0$ and/or $\Delta\mC_{2}\neq 0$, we call that scenario as ``$\widetilde{\rm SP}$''. In SP and $\widetilde{\rm SP}$ scenarios, the NP WCs corresponding to the vector operators are taken to be zero.

In \cite{Bobeth:2011st}, current bounds on the scalar operators are calculated indirectly from $b\rightarrow s \gamma$, $b\rightarrow s l^+ l^-$ processes and directly from  $B_s\rightarrow \tau^+\tau^-$, $B^+\rightarrow K^+\tau^+\tau^-$ and  $B^+\rightarrow X_s\tau^+\tau^-$ processes. These bounds are rather weak:
\begin{align}
    |\mC_S\pm \mC_P| &< 0.4\frac{2\pi}{\alpha_e}~,\quad
    |\mC_S^{\prime}\pm \mC_P^{\prime}| < 0.4\frac{2\pi}{\alpha_e}~.
\end{align} 
%

\begin{figure*}[t]
    \centering
    \includegraphics[width=0.31\textwidth]{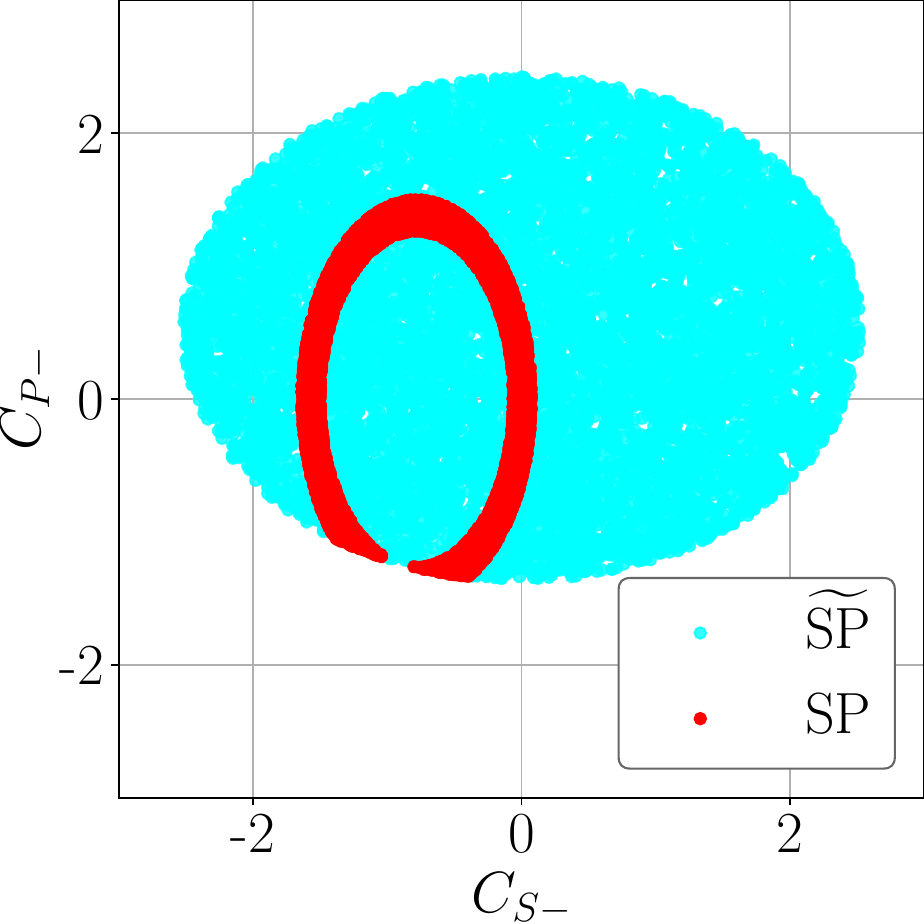}
    \qquad\includegraphics[width=0.31\textwidth]{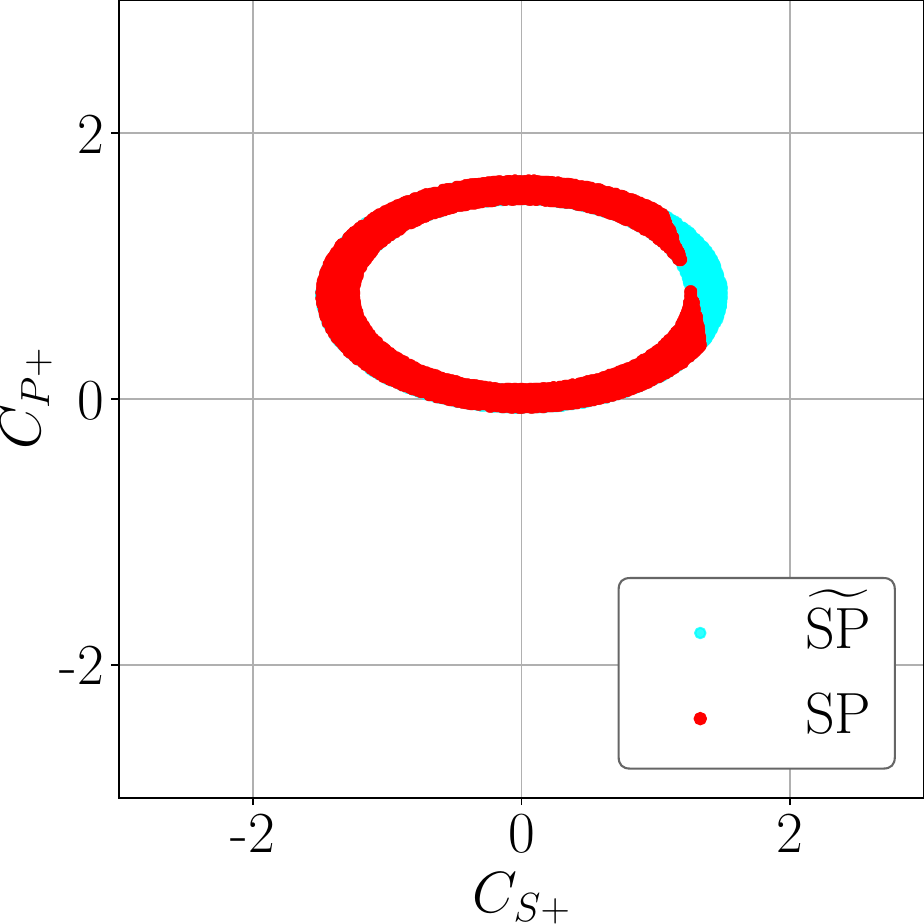}
    \caption{\justifying Projected bounds on the NP scalar WCs in the ($C_{S-}$, $C_{P-}$) and ($C_{S+}$, $C_{P+}$) planes,  from the expected upper bound on  ${\cal B}(B_s\to \tau \tau) < 10^{-5}$ and the expected measurement of ${\cal B}(B^+\to K^+ \tau \tau)$ with 10\% of its SM value. Note that the SP regions (red) are subsets of $\widetilde{\rm SP}$ region (cyan), i.e. all the red regions have cyan regions underneath.}\label{fig:SPregion}
\end{figure*}

In our analysis, we take $\mathcal{B}(B_s \rightarrow \tau^+ \tau^-)<10^{-5}$, which is the expected upper bound at the HL-LHC and FCC-ee~\cite{FCC:2018byv}, and  the expected measurement of ${\cal B}(B\to K\tau^+\tau^-)$ at FCC-ee to be within the precision of $10\%$ of the SM value. With these assumptions, we find the allowed region for the four parameters $C_{S+}$, $C_{P+}$, $C_{S-}$ and $C_{P-}$ for the two scenarios SP and $\widetilde{\rm SP}$. We present the projections of these allowed regions in Fig.~\ref{fig:SPregion}.
Note that the observable ${\cal B}(B\to K\tau^+\tau^-)$ is more sensitive to $C_{S+}$ and $C_{P+}$, whereas $\mathcal{B}(B_s \rightarrow \tau^+ \tau^-)$ is more sensitive to $C_{S-}$ and $C_{P-}$. The upper bound $\mathcal{B}(B_s \rightarrow \tau^+ \tau^-) < 10^{-5}$ constrains $C_{S-}$ and $C_{P-}$ to a circular cyan region. On the other hand, the constraint on ${\cal B}(B\to K\tau^+\tau^-)$ leads to an annular red region in the ($C_{S+}$, $C_{P+}$) plane. 

Note that in Fig.~\ref{fig:VAregion} and Fig.~\ref{fig:SPregion}, we have not accounted for uncertainties due to form factors and other input parameters. When these uncertainties are included, the allowed regions broaden, although their overall shapes remain largely unchanged. We shall later include the uncertainties and study their effects.

Based on the allowed WCs values, our next goal is to find observables whose values for the $\widetilde{\rm SP}$ scenario will be distinguishable from those for the VA and SP scenarios.

\section{Angular observables in  $B^0\to K^{*0}\tau^+ \tau^-$}\label{sec:results}

In this section, we study the effects of the  $\widetilde{\textrm{SP}}$ scenario, in comparison to the SM, VA and SP scenarios, on the angular observables in $B\to V(\to P_1\,P_2) \tau^+ \tau^-$ decays. Here $V$ and $P_{1(2)}$ are vector and pseudoscalar mesons, respectively. The angular distribution for $B\to V (\to P_1\,P_2) \tau^+ \tau^-$ decay is given as~\cite{Altmannshofer:2008dz}
\begin{equation}\label{eq:d4Gamma}
    \frac{d^4\Gamma}{dq^2\, d\cos\theta_l\, d\cos\theta_{V}\, d\phi} =
    \frac{9}{32\pi} I(q^2, \theta_l, \theta_{V}, \phi)\,,
\end{equation}
where
\begin{widetext}
\begin{align} \label{eq:angulardist}
    I(q^2, \thl, \thV, \phi)& = I_1^s \sin^2\thV + I_1^c \cos^2\thV
    + (I_2^s \sin^2\thV + I_2^c \cos^2\thV) \cos 2\thl + I_3 \sin^2\thV \sin^2\thl \cos 2\phi \nonumber \\       
    & + I_4 \sin 2\thV \sin 2\thl \cos\phi  + I_5 \sin 2\thV \sin\thl \cos\phi + (I_6^s \sin^2\thV + {I_6^c \cos^2\thV})\cos\thl 
    \nonumber \\           
    & + I_7 \sin 2\thV \sin\thl \sin\phi + I_8 \sin 2\thV \sin 2\thl \sin\phi
    + I_9 \sin^2\thV \sin^2\thl \sin 2\phi\,.
\end{align}
\end{widetext}
Here the direction of $V$ in the rest frame of B is taken to be the $z$-axis. The angle between the $z$ axis and the direction of $P_1$ in the rest frame of V is denoted by $\thV$. The angle between the $-z$ axis and the direction of $\tau^-$ in the center-of-mass frame of the lepton pair is denoted by $\thl$. The azimuthal angle $\phi$ is the angle between the decay planes formed by the decay products $(P_1, P_2)$ and $(\tau^+, \tau^-)$ in the rest frame of $B$. In our analysis, $V \equiv K^{*0}$, $P_1\equiv K^+$ and  $P_2 \equiv \pi^-$.

We calculate the values of the following observables as functions of $q^2$ using the software package {\tt flavio}~\cite{Straub:2018kue}:
\begin{align}
    S_i^{(a)} & = \frac{(I_i^{(a)} + \bar I_{i}^{(a)})}{d(\Gamma + \bar\Gamma)/dq^2}\,,~~ A_{FB} = \frac{3}{8}(2S_6^s + S_6^c)\,,~~ F_L = S_1^c,
\end{align}
where $\bar I_{i}^{(a)}$ and $\bar \Gamma$ are the corresponding angular coefficient and decay width for the CP-conjugate process. The explicit expressions of these angular observables are available in \cite{Altmannshofer:2008dz}. In Table~\ref{tab:sensitivity}, we present the observables that are sensitive to each NP parameter.

\begin{figure}
    \raggedright
    \quad\includegraphics[width=0.36\textwidth]{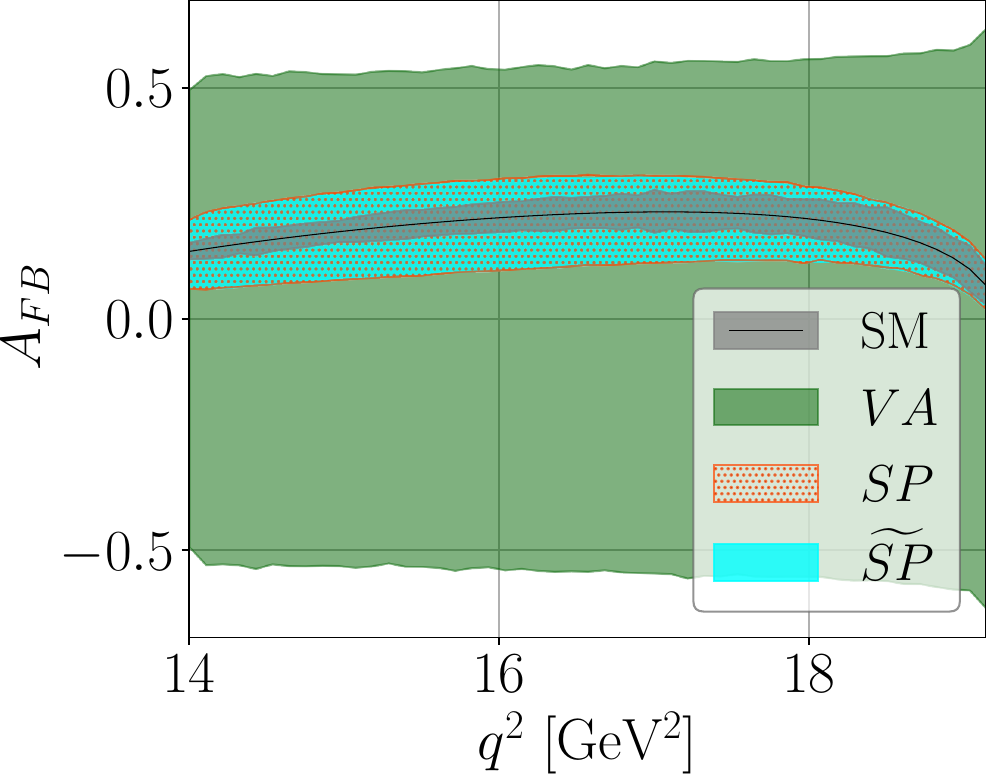}\\
    \vspace{0.2cm}
    \quad\includegraphics[width=0.36\textwidth]{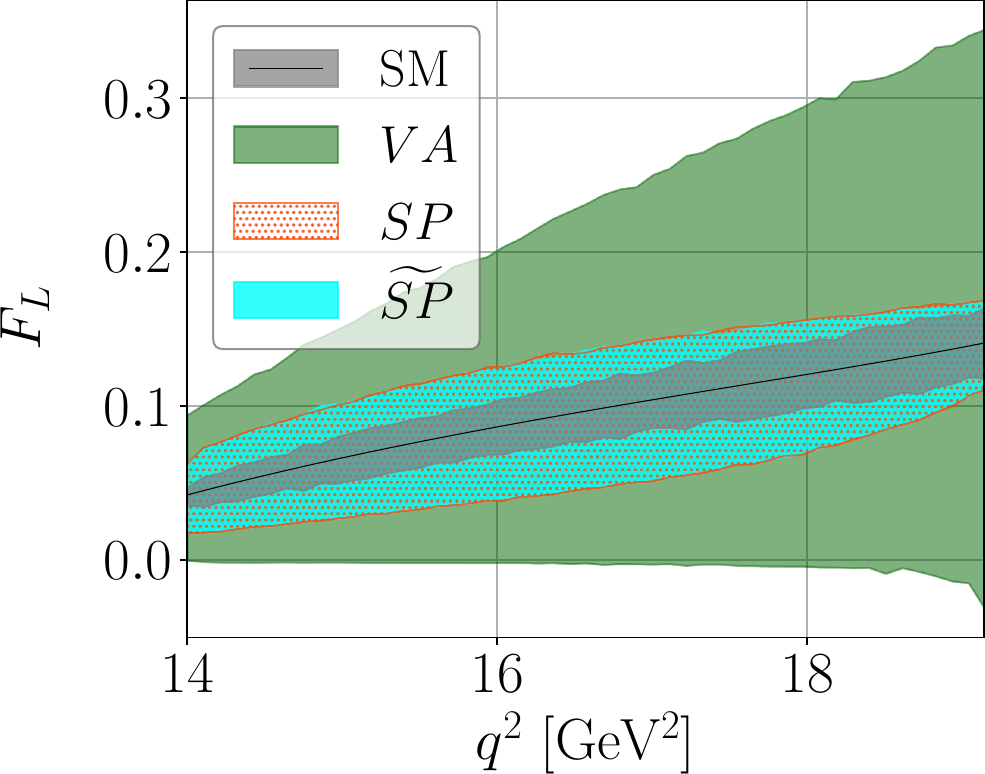}\\
    \vspace{0.2cm}
    \quad\includegraphics[width=0.36\textwidth]{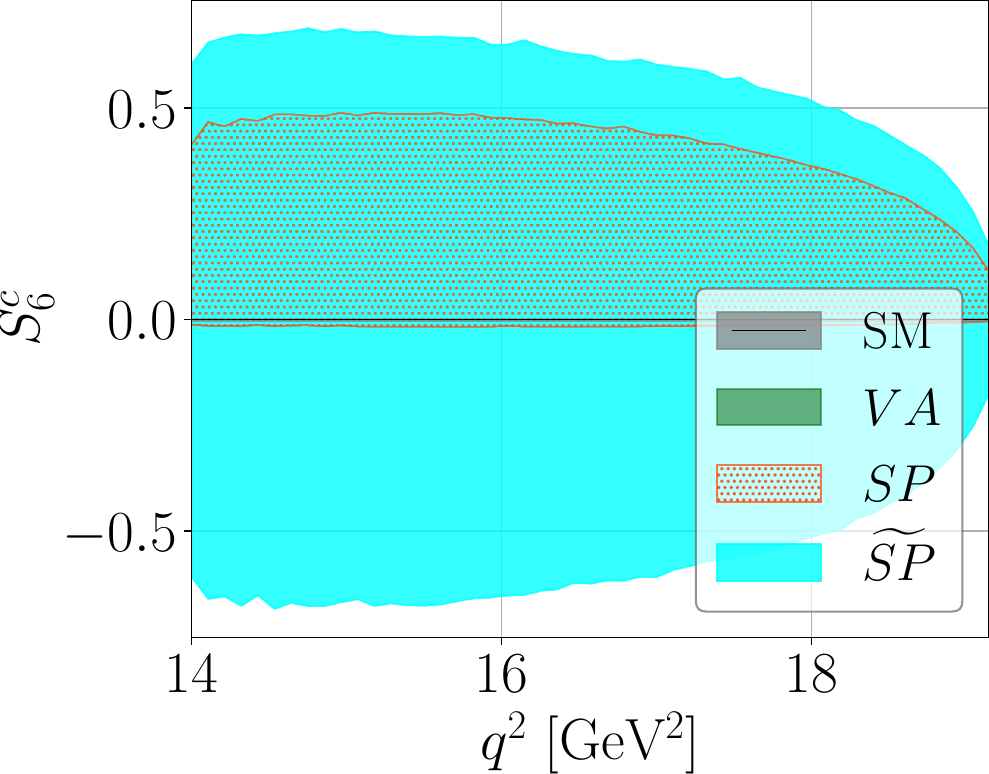}
    \caption{\justifying The allowed values of angular observables  $A_{FB}$ (top panel) $F_L$ (middle panel) and $S_6^c$ (bottom panel) in $B^{0}\rightarrow K^{0*} \tau^+\tau^-$ in the scenarios SM, VA, SP and $\widetilde{\rm SP}$ with $2\sigma$ theoretical uncertainties included.}
    \label{fig:BtoK*0}
\end{figure}	

We calculate these observables in $B^0\to K^{*0}\tau^+\tau^-$ for the SM and the NP scenarios VA, SP and $\widetilde{\rm SP}$. Note that when we calculate the expected values of the observables in the VA scenario, we consider points for $C_9^{(\prime)}$ and $C_{10}^{(\prime)}$ from the allowed regions as shown in Fig.~\ref{fig:VAregion}, while keeping all WCs for the scalar operators as zero. Similarly, while calculating observables in SP ($\widetilde{\rm SP}$) scenario, we put $C_9^{(\prime)} = C_{10}^{(\prime)} =0$ and take $C_{S,P}^{(\prime)}$ points from the red (cyan) regions shown in Fig.~\ref{fig:SPregion}.  
We find that the angular observables $A_{FB}$, $F_L$, and $S_6^c$ show significant sensitivity to the SP and $\widetilde{\rm SP}$ scenarios. In Fig.~\ref{fig:BtoK*0}, we present the expected values of $A_{FB}$, $F_L$ and $S_6^c$ in SM and in the NP scenarios VA, SP and $\widetilde{\rm SP}$. For all these four observables, we have considered $2\sigma$ uncertainties due to form factors and other input parameters, and the WCs corresponding to the NP scenarios are varied within their allowed regions. 

\begin{table}[t]
	\centering
	\begin{tabular}{|c|c|}
            \hline
            NP WCs & Sensitive observables\\
            \hline
            $C_9^{(\prime)}$, $C_{10}^{(\prime)}$ & \makecell*{$S_1^{s,c}$, $S_2^{s,c}$, $S_3$, $S_4$, $S_5$, $S_6^s$, $A_{7}$\\
            $A_{FB}$, ${\cal B}(B\to K^{*}\tau^+\tau^-)$}\\
            \hline
            $C_{S-}$ & \makecell*{$S_1^c+S_2^c$, $S_6^c$, $A_{FB}$ \\ ${\cal B}(B_s\to\tau^+\tau^-)$}\\
            \hline
            $C_{P-}$ & \makecell*{$F_L$ \\ ${\cal B}(B_s\to\tau^+\tau^-)$}\\
            \hline
            $C_{S+}$, $C_{P+}$ & \makecell*{ ${\cal B}(B\to K\tau^+\tau^-)$}\\
            \hline
	\end{tabular}
	\caption{\justifying Wilson coefficients of NP vector and scalar operators, and observables sensitive to each of them.}\label{tab:sensitivity}
\end{table} 

As shown in Fig.~\ref{fig:BtoK*0}, the observables $A_{FB}$ and $F_L$ by themselves would not be able to distinguish the effects of $\widetilde{\rm SP}$ from VA and SP scenarios or even from the SM. However, the observable $S_6^c$ is suppressed for both SM and VA and may have significant nonzero values for SP and $\widetilde{\rm SP}$ scenarios. Moreover, some values of $S_6^c$ that are possible in $\widetilde{\rm SP}$ are not achievable in SP scenario. Thus $S_6^c$ is a good probe for distinguishing between scenarios within SMEFT (VA and SP) and beyond SMEFT ($\widetilde{\rm SP}$). For example,  even a small negative value of $S_6^c$ will hint towards physics beyond SMEFT.

\begin{figure*}[t]
    \centering
    \includegraphics[width=0.31\textwidth]{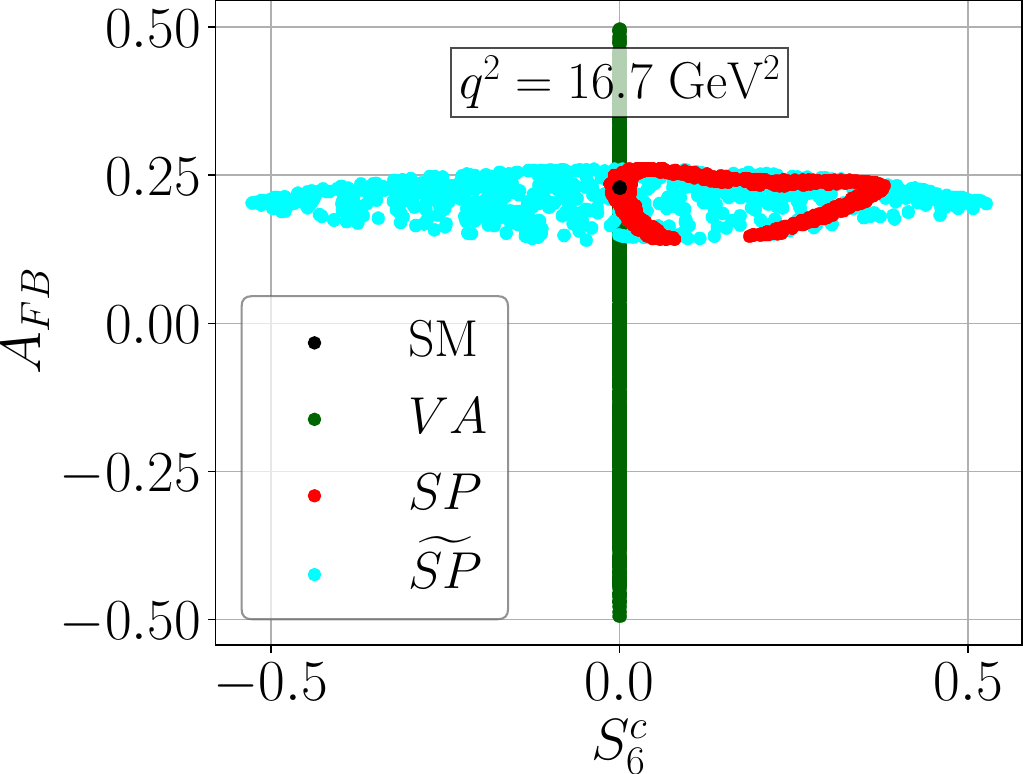}
    \quad\includegraphics[width=0.31\textwidth]{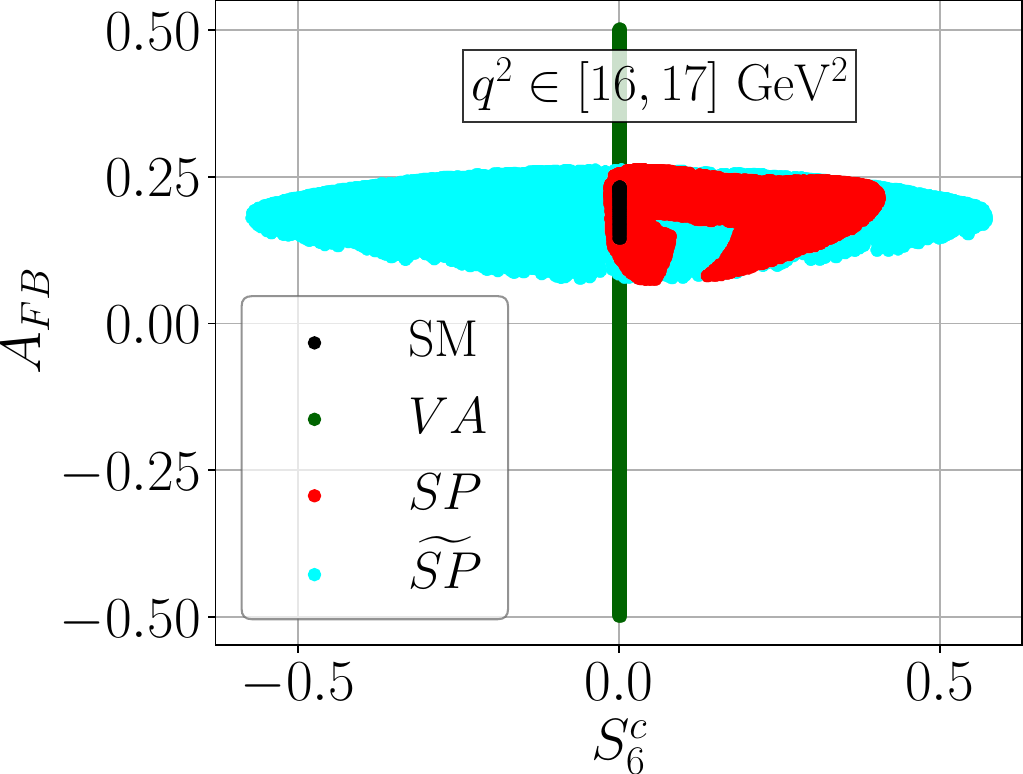}
    \quad\includegraphics[width=0.31\textwidth]{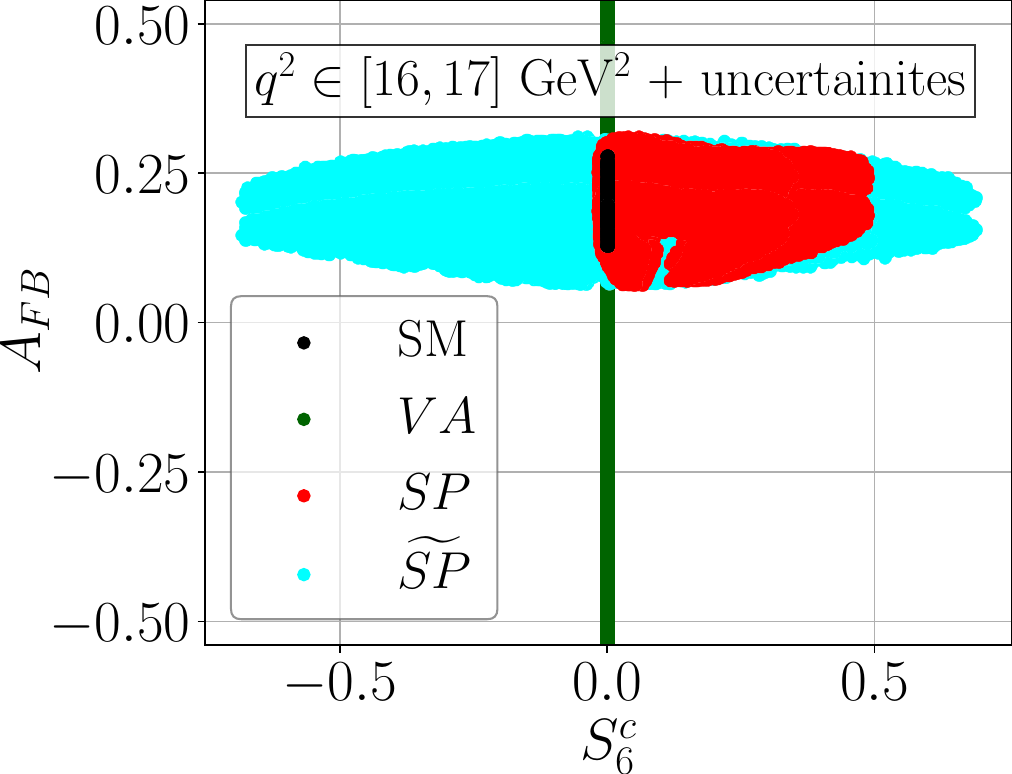}\\
    \vspace{0.2cm}
    \includegraphics[width=0.31\textwidth]{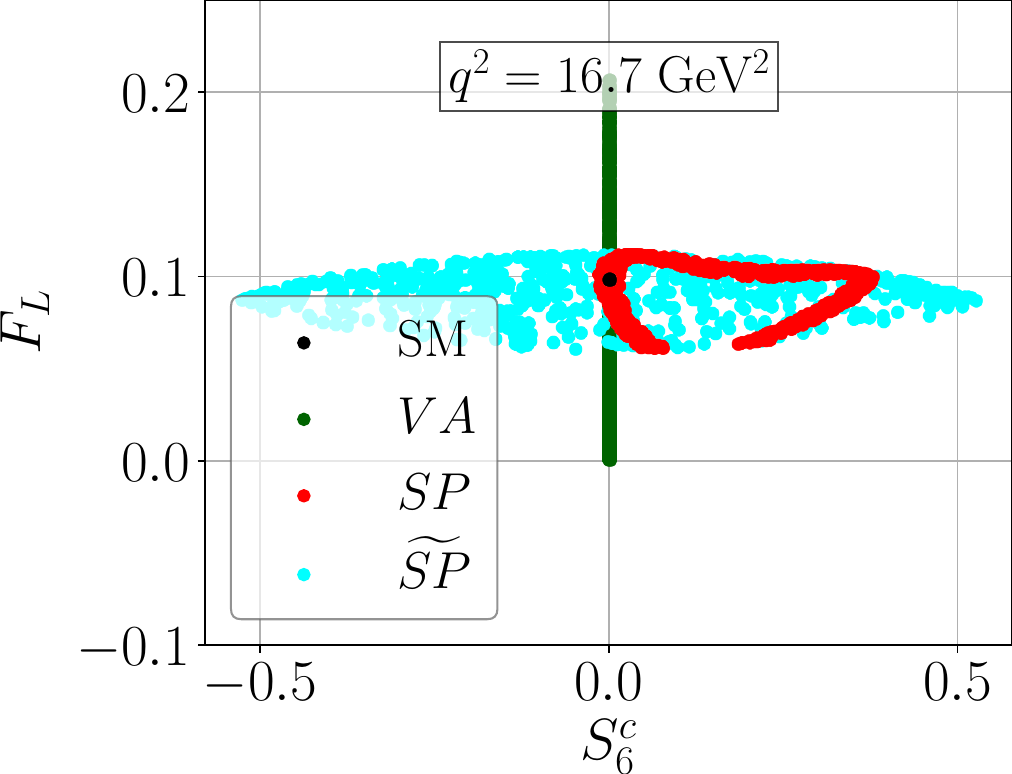}
    \quad\includegraphics[width=0.31\textwidth]{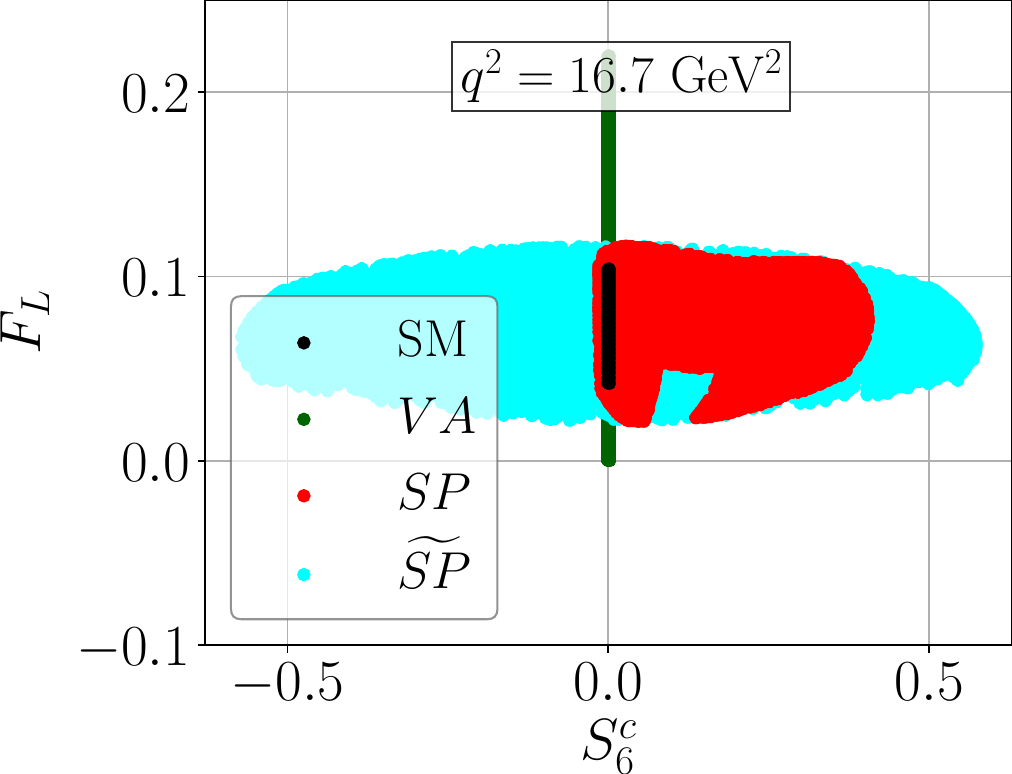}
    \quad\includegraphics[width=0.31\textwidth]{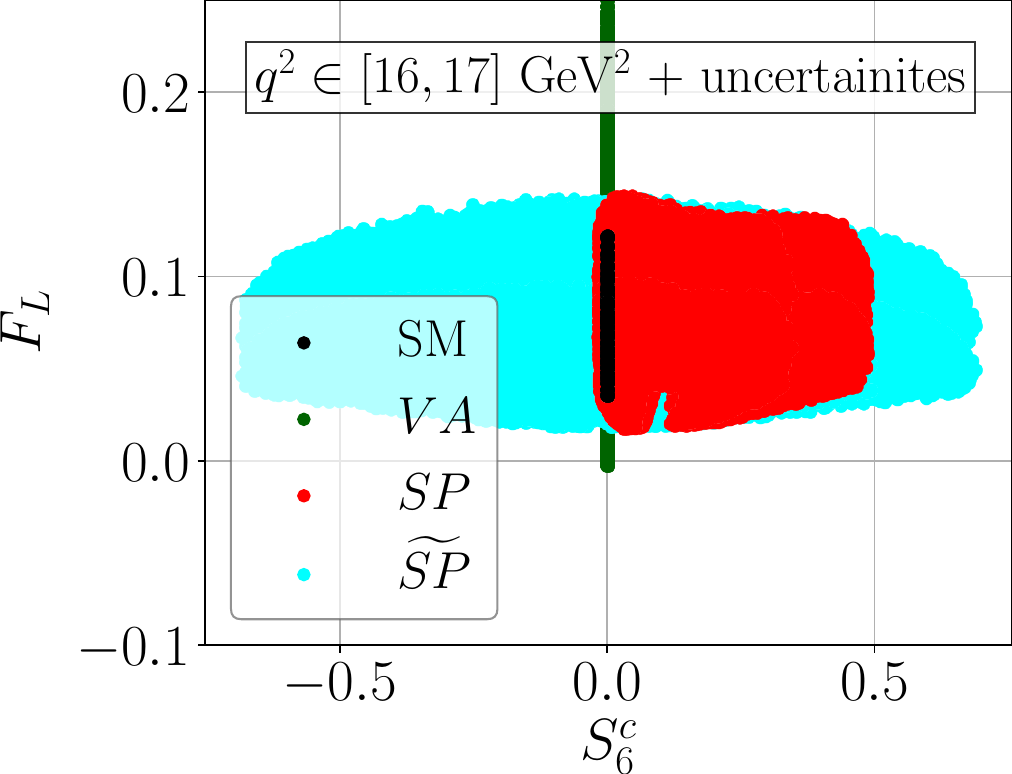}
    \caption{\justifying The combinations of the pairs of angular observables ($A_{FB}$, $S_6^c$) and ($F_L$, $S_6^c$) in $B^{\pm}\rightarrow K^{0*} \tau^+\tau^-$, for the SM and for the three NP scenarios VA, SP and $\widetilde{\rm SP}$. The left column corresponds to the fixed benchmark value $q^2 = 16.7$ GeV$^2$. The middle column is for the bin $q^2 \in [16,17]$ GeV$^2$. The right column is for the same $q^2$ bin, but after including the $2\sigma$ uncertainties due to form factors and other input parameters. }
    \label{fig:q2binRealNoErr}
\end{figure*}	

Next, we ask whether it is possible to distinguish the different scenarios even in the case of $S_6^c$ having a positive value. To answer this, we consider the pairs of angular observables  ($A_{FB}$, $S_6^c$) and ($F_{L}$, $S_6^c$). 
The left panels of Fig. \ref{fig:q2binRealNoErr} show the possible points in the respective planes for a fixed value of $q^2$ and in the absence of any uncertainties in the input parameters. This indicates that in principle the combination $S_6^c$ with another observable would be more efficient than $S_6^c$ alone in identifying the beyond-SMEFT effects. However, the necessity of binning over $q^2$ reduces the efficacy of such combinations as shown in the central panels of Fig.~\ref{fig:q2binRealNoErr}. The inclusion of $2\sigma$ uncertainties due to form factors and input parameters further limits the ability of these combinations of observables as probes of physics beyond SMEFT. This may be seen from the right panels of Fig.~\ref{fig:q2binRealNoErr}. Thus, one would need narrow $q^2$ bins (hence a large number of events) and better control over theory parameters in order to effectively exploit such combinations.

\section{Results with complex NP Wilson coefficients}\label{sec:complex}

In our analysis so far, we have considered all the WCs to be real. However, the WCs in LEFT can be complex in general. Note that even if all the WCs in SMEFT (or HEFT) in the UV scale are real in the flavor basis, phases will appear in low-energy WCs through CKM elements while matching. The basic idea and results of our analysis are independent of whether the WCs are real or complex. However, it is worthwhile to see the effects of complex WCs on the angular observables, especially the ones that are asymmetric under charge-parity (CP) transformation.  

\begin{figure}
    \centering
    \includegraphics[width=0.22\textwidth]{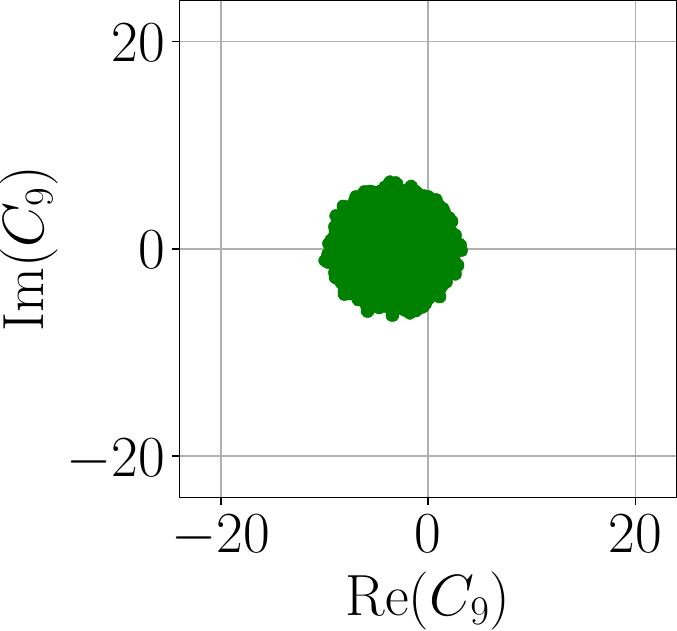}
    \quad\includegraphics[width=0.22\textwidth]{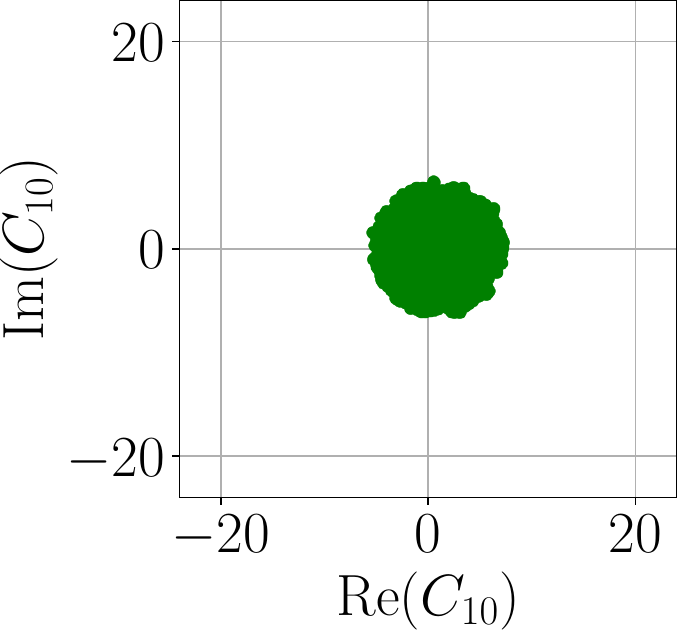}\\
    \bigskip\includegraphics[width=0.22\textwidth]{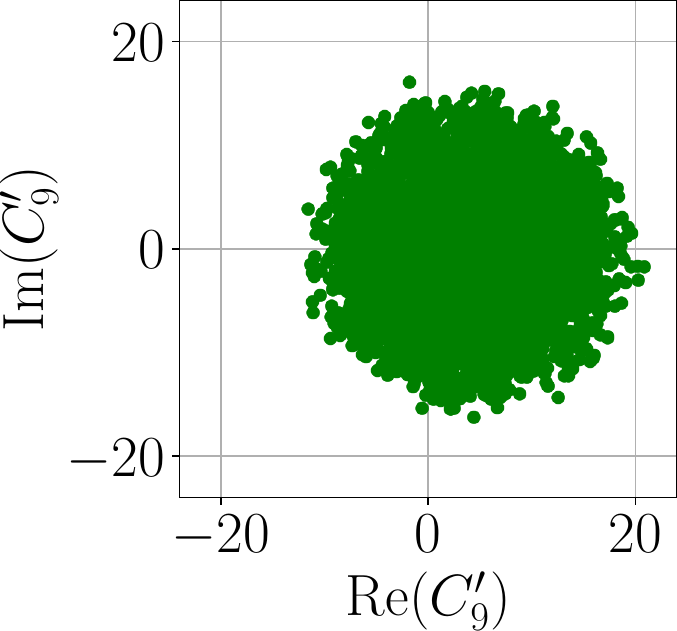}
    \quad\includegraphics[width=0.22\textwidth]{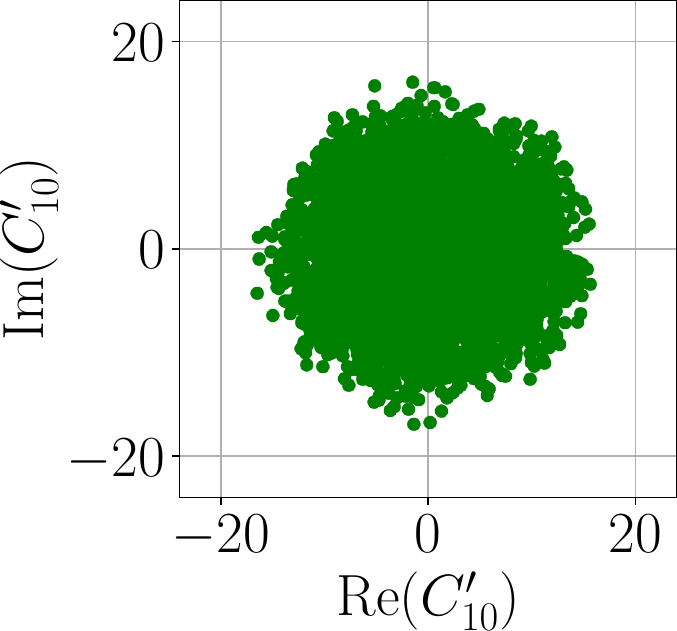}
    \caption{\justifying Projected bounds on the complex parameters $C_9$, $C_{10}$, $C_{9}^\prime$ and $C_{10}^\prime$ from the expected observation of ${\cal B}(B\to K^{*}\tau\tau)$ at FCC-ee to be within 10\% of its SM value.}
    \label{fig:VAcomplex}
\end{figure}

In this section, we consider the WCs $\mC_{9}^{(\prime)}$, $\mC_{10}^{(\prime)}$, $\mC_{S\pm}$ and $\mC_{P\pm}$ to be complex.  In Fig.~\ref{fig:VAcomplex}, we present the allowed regions for complex $\mC_{9}^{(\prime)}$ and $\mC_{10}^{(\prime)}$ in the VA scenario. The methodology for obtaining these constraints is the same as discussed in Sec.~\ref{sec: bounds}. Note that we allow all of these parameters to be nonzero at the same time, however, we do not show the correlations among them in Fig.~\ref{fig:VAcomplex}. The regions are symmetric about the real axis as expected. 

\begin{figure}
    \centering
    \includegraphics[width=0.21\textwidth]{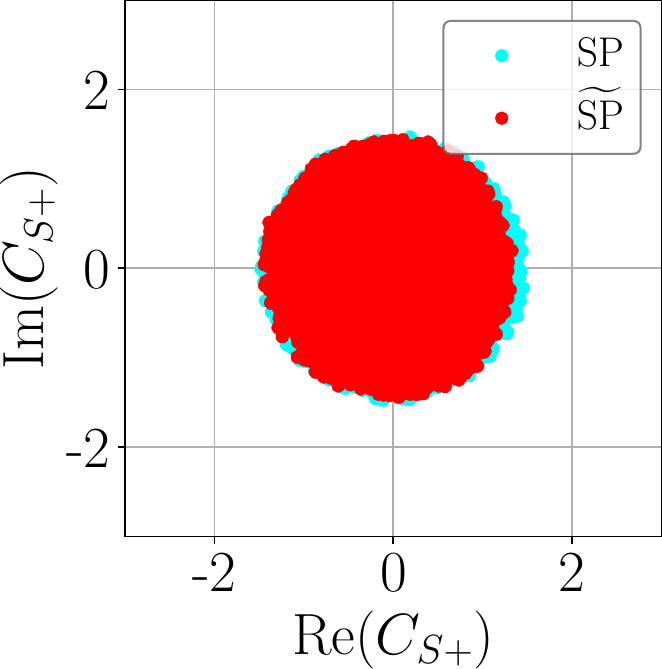}
    \quad\includegraphics[width=0.22\textwidth]{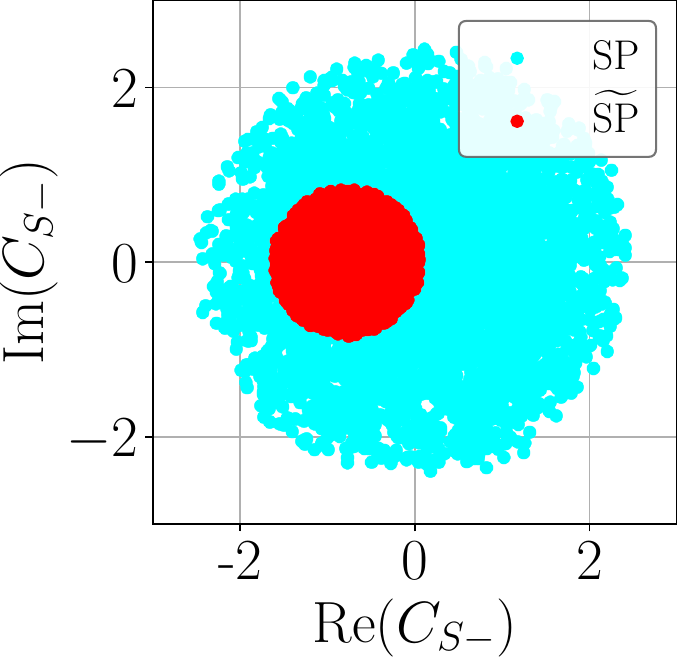}\\
    \vspace{0.1cm}
    \includegraphics[width=0.21\textwidth]{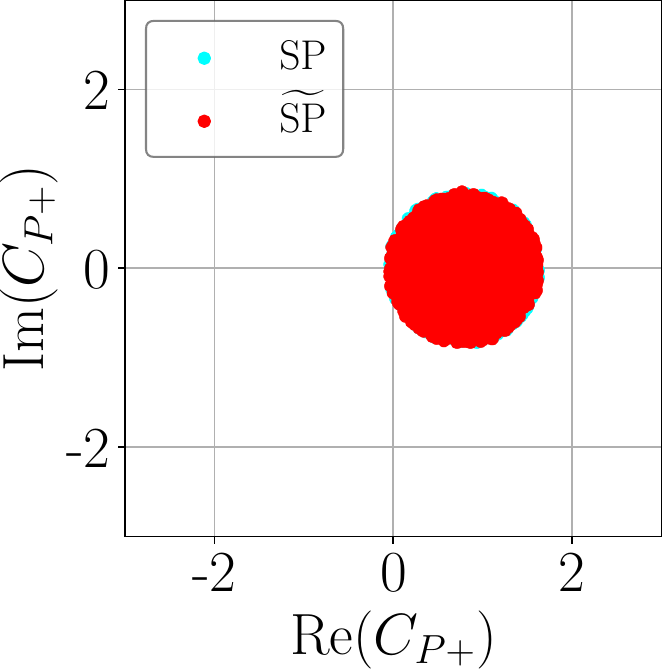}
    \quad\includegraphics[width=0.22\textwidth]{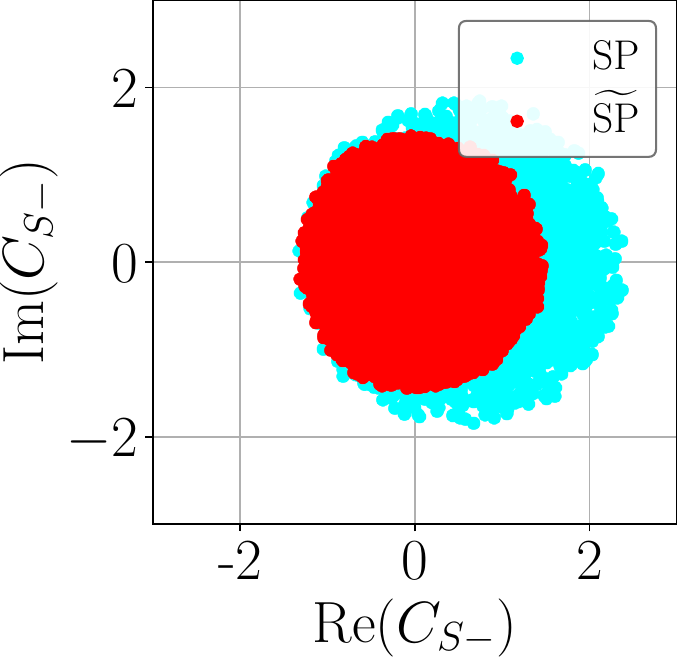}
    \caption{\justifying Projected bounds for the complex parameters $C_{S-}$, $C_{P-}$, $C_{S+}$, and $C_{P+}$ from the expected upper bound on  ${\cal B}(B_s\to \tau \tau)$ and the expected measurement of ${\cal B}(B^+\to K^+ \tau \tau)$ at HL-LHC/FCC-ee. Note that the SP regions (red) are subsets of $\widetilde{\rm SP}$ region (cyan), i.e. all the red regions have cyan regions underneath.}
    \label{fig:SPcomplex}
\end{figure}

In Fig.~\ref{fig:SPcomplex}, we show the allowed regions for complex $\mC_{S\pm}$ and $\mC_{P\pm}$ in the SP and $\widetilde{\rm SP}$ scenarios. It is observed that the allowed regions in $C_{S+}$ and $C_{P+}$ parameter spaces in the two scenarios SP and $\widetilde{\rm SP}$  overlap almost completely. However, for $C_{S-}$ and $C_{P-}$, there are regions in this complex parameter space that are only accessible in the scenario beyond SMEFT ($\widetilde{\rm SP}$).

\begin{figure}[t]
    \centering
    \includegraphics[width=0.38\textwidth]{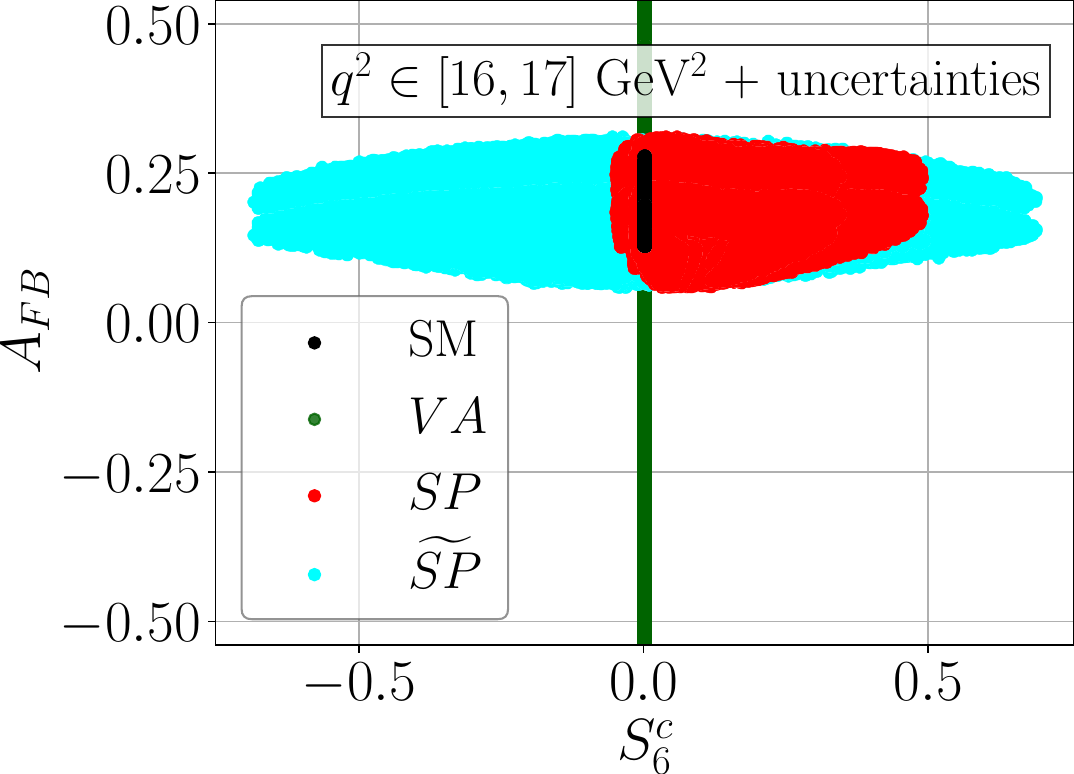}\\
    \vspace{0.2cm}
    \includegraphics[width=0.38\textwidth]{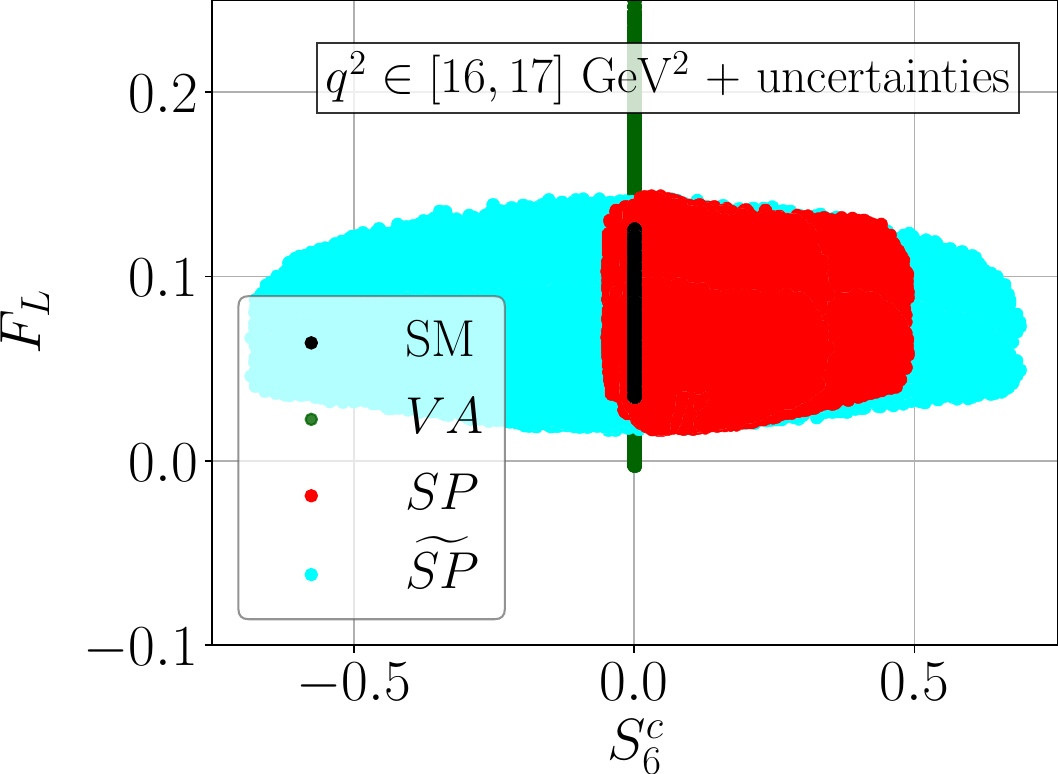}
    \caption{\justifying The combinations of the pairs of angular observables ($A_{FB}$, $S_6^c$) and ($F_L$, $S_6^c$) in $B^{\pm}\rightarrow K^{0*} \tau^+\tau^-$, for the SM and for the three NP scenarios VA, SP and $\widetilde{\rm SP}$ with complex NP parameters. These plots are for the bin  $q^2 \in [16,17]$ GeV$^2$ with the inclusion of $2\sigma$ theoretical uncertainties.}
    \label{fig:obscombcomplex}
\end{figure}

The results for the observables $A_{FB}$, $F_L$, and $S_{6}^c$ are similar to the ones discussed in Sec.~\ref{sec:results} with allowed regions becoming wider. We do not show the results with these individual observables in this section. The results for the combinations ($A_{FB}$, $S_{6}^c$) and ($F_L$,, $S_{6}^c$) are shown in Fig.~\ref{fig:obscombcomplex}.  These indicate that for complex WCs, a significantly negative value of $S_6^c$ ($S_6^c \lesssim -0.2$) as well as a large positive value of $S_6^c$  ($S_6^c \gtrsim 0.5$) would suggest physics beyond SMEFT. The observables $A_{FB}$ and $F_L$ do not seem to be offering any extra advantage to identify $\widetilde{\rm SP}$. 

\begin{figure}[t]
    \centering
    \includegraphics[width=0.38\textwidth]{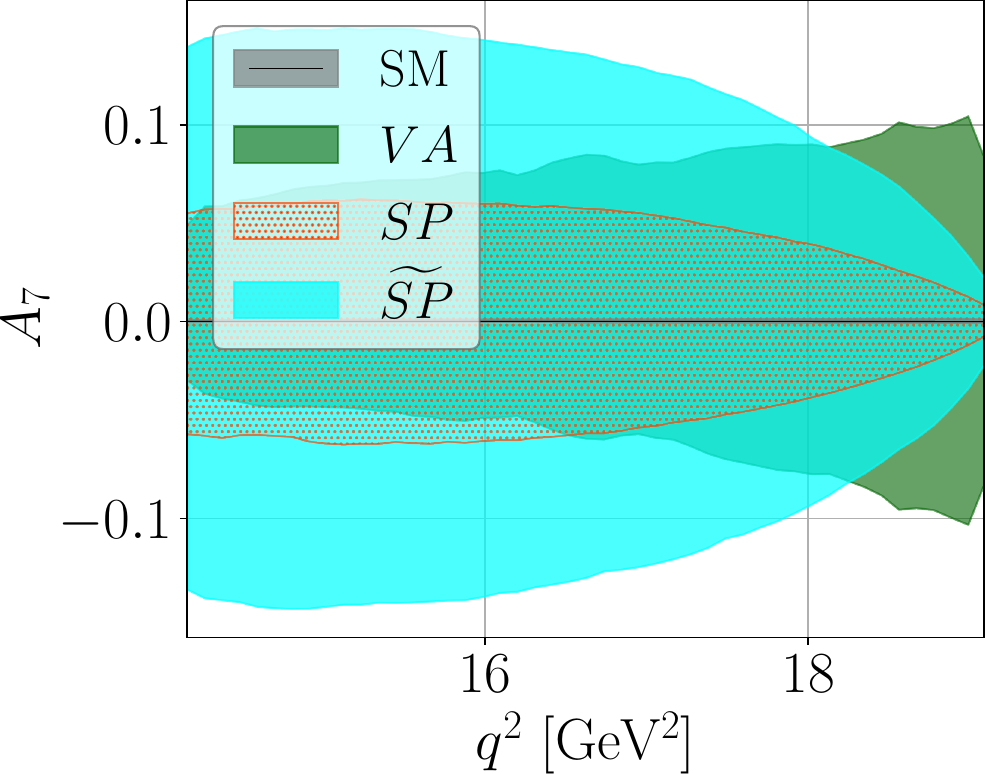}\\
    \vspace{0.2cm}
    \includegraphics[width=0.38\textwidth]{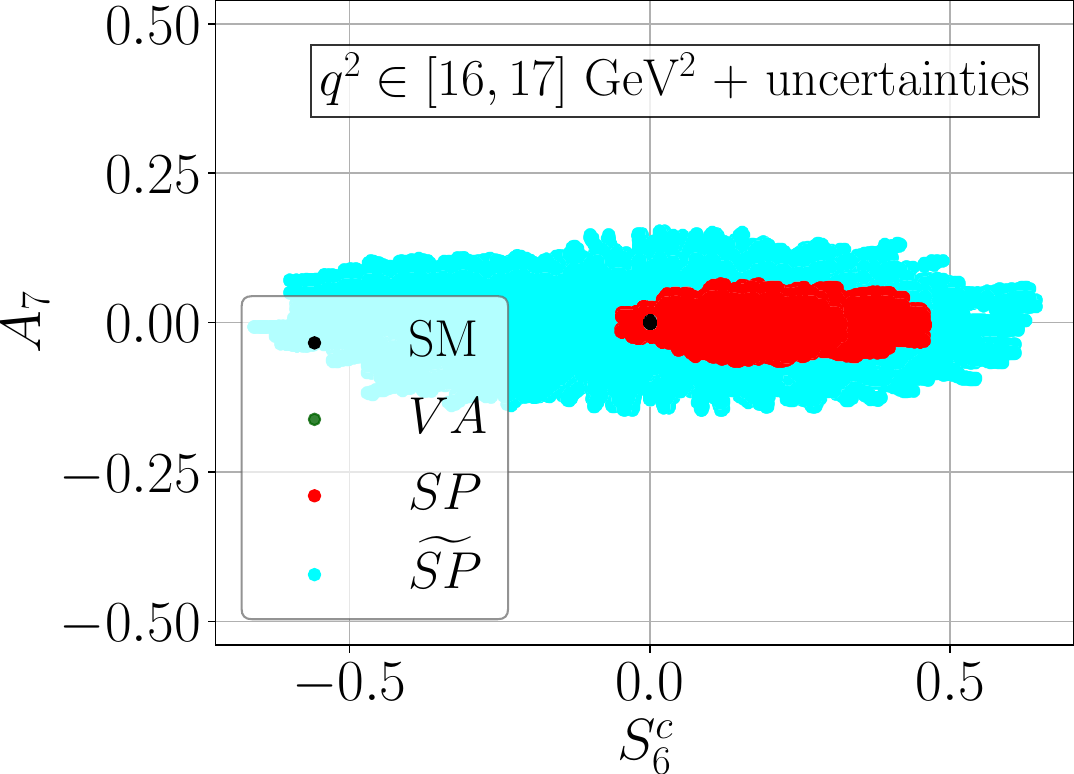}
    \caption{\justifying Allowed values of $A_7$ as a function of $q^2$ (top), and the combination of $A_7$ and $S_{6}^c$ in the bin $q^2\in [16, 17]$ GeV$^2$ (bottom) for complex NP WCs, with $2\sigma$ theoretical uncertainties included.}
    \label{fig:A7}
\end{figure}

We further explore the CP asymmetric observable $A_7$ defined as 
\begin{align}
    A_7 & = \frac{I_7 - \bar I_7}{d(\Gamma + \bar \Gamma)/dq^2}~.\label{A7def}
\end{align}
In the SM as well as in the NP scenarios where all the WCs are real,  $A_7$ is highly suppressed. However, for complex WCs, it may become significantly nonzero. The value of this observable as a function of $q^2$ is shown in the top panel of Fig.~\ref{fig:A7}. Clearly, $A_7$ has the ability to identify the $\widetilde{\rm SP}$ scenario by itself. The combination of $A_7$ and $S_6^c$ can be more effective at this task than the two observables separately, even after taking into account the $q^2$ binning of $1$ GeV$^2$ and $2\sigma$ theoretical uncertainties, as can be seen in the bottom panel of  Fig.~\ref{fig:A7}.

\section{Concluding remarks}\label{sec:conclusion}

The Standard Model effective field theory (SMEFT) is often taken to be the default EFT above the EW scale. However, more general EFTs, such as Higgs effective field theory (HEFT) where the EW gauge symmetry $SU(2)_L\times U(1)_Y$ is non-linearly realized, are still not ruled out. In this paper, we study the possibility of probing beyond-SMEFT physics in low-energy processes mediated via the $b\to s \tau\tau$ transition. 

When relevant operators in the low-energy effective field theory (LEFT) are mapped to the EFT above the EW scale, certain relations that are satisfied in SMEFT may not hold if a more general EFT such as HEFT is valid over the EW scale. 
For the neutral-current transition $b\to s \tau\tau$, some such relations are $C_S = -C_P$, $C_S^\prime = C_P^\prime$, which are predicted by SMEFT. We identify observables that are sensitive to any deviations from these relations and explore the possibility of distinguishing these deviations from not only SM, but also from other NP scenarios that are still within SMEFT. 

We consider three NP scenarios (i) VA, where only NP vector operators are nonzero, (ii) SP, where only NP scalar operators are nonzero but SMEFT-predicted relations are obeyed, and (iii) $\widetilde{\rm SP}$, where only NP scalar operators are nonzero and the SMEFT-predicted relations are not imposed. We calculate constraints on the NP WCs in these three scenarios using projected observations of ${\cal B}(B_s\to \tau^+\tau^-)$ and ${\cal B}(B\to K^{(*)} \tau^+\tau^-)$ with the precision expected in HL-LHC and the phase-I of FCC-ee. We take these measurements to be consistent with the SM; specifically, we consider an upper bound ${\cal B}(B_s\to \tau^+\tau^-) < 10^{-5}$ and a measurement of ${\cal B}(B\to K^{(*)} \tau^+\tau^-)$ within 10\% of its SM prediction.

We calculate the values of angular observables in  $B\to K^{0*} \tau^+\tau^-$, as functions of $q^2$, in the SM and the three NP scenarios mentioned above, possible with the allowed real values of the NP WCs. We find that the angular observable $S_6^c$ will be the most efficient in distinguishing the beyond-SMEFT scenario $\widetilde{\rm SP}$ from the SM as well as from the other NP scenarios. A significant negative value of $S_{6}^c$ $(\lesssim - 0.2)$ or a large positive value of $S_6^c$ $(\gtrsim  0.5)$ would suggest physics beyond SMEFT. The observables $A_{FB}$ and $F_L$ can help in this identification in principle. However, with a 1 GeV$^2$ binning in $q^2$ and the inclusion of current theoretical uncertainties, the usefulness of these observables would seem rather constrained. 

We further consider the possibility that the NP WCs can take complex values and repeat the above analysis. This allows us to access an additional angular observable $A_7$ which is sensitive to beyond-SMEFT effects. Moreover, the combination of $A_7$ and $S_6^c$ would be more effective compared to the two observables by themselves. Thus, we demonstrate that the NP that would be hidden if we only looked at the branching ratios could manifest itself in these angular observables.

Although the measurements of the two branching ratios considered in our analysis will only be available after the full run of HL-LHC (for $B_s\to \tau^+\tau^-$) and phase-I of FCC-ee (for $B\to K^{(*)} \tau^+\tau^- $), the analysis proposed is a promising way to identify scenarios beyond SMEFT through neutral-current semileptonic decays of $B$ mesons and underscores the importance of efficient $\tau$ detection in future collider experiments.

\section*{Acknowledgements}

We would like to thank Susobhan Chattopadhyay, Rick S. Gupta, Gagan B. Mohanty and Tuhin S. Roy for useful discussions. This work is supported by the Department of Atomic Energy, Government of India, under Project Identification Number RTI 4002.  We acknowledge the use of computational facilities of the Department of Theoretical Physics at Tata Institute of Fundamental Research, Mumbai. We would also like to thank Ajay Salve and  Kapil Ghadiali for technical assistance. 

\begin{figure}[t!]
    \centering
    \includegraphics[width=0.34\textwidth]{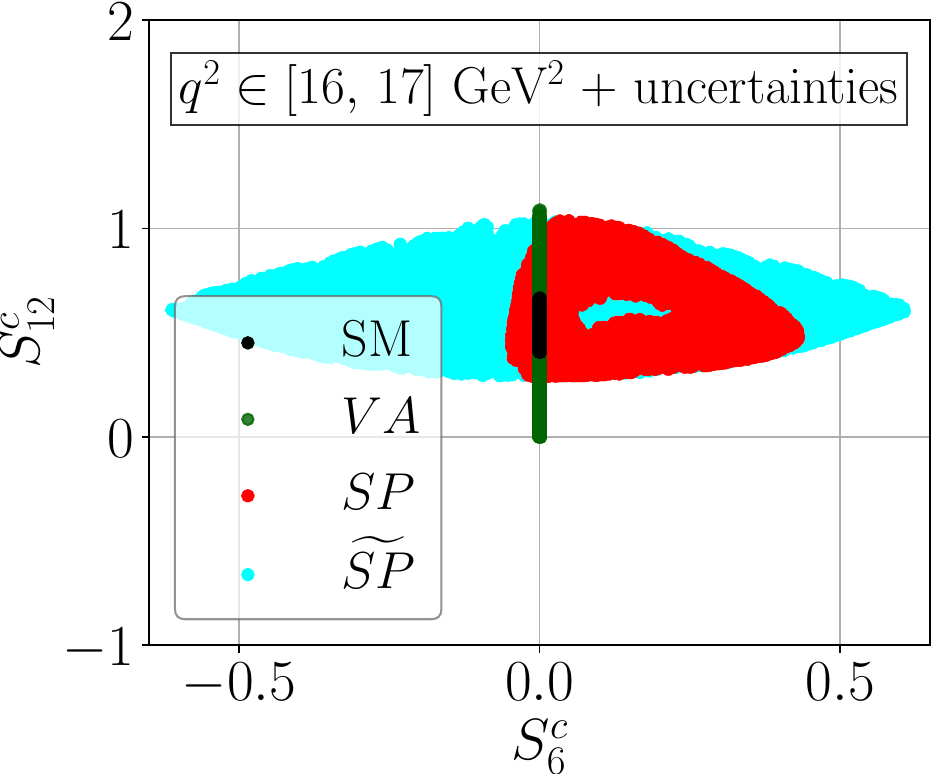}
    \caption{\justifying Allowed values of the pair of observables ($S_{12}^c$, $S_6^c$) for the bin $q^2\in [16, 17]$ GeV$^2$ with $2\sigma $ theoretical uncertainties included. }
    \label{fig:S12combs}
\end{figure}

\appendix
\section{The angular observable $S_1^c + S_2^c$}\label{sec:S12c}
In Table~\ref{tab:sensitivity} we have mentioned that the observable $S_{12}^c \equiv S_1^c + S_2^c$ is sensitive to $C_{S-}$. Indeed, it gets a contribution from $|C_{S-}|^2$ at the leading order while the contributions from $C_{P-}$ and the vector WCs are suppressed by a factor of $m_\tau^2/q^2\sim 1/5$.
In this appendix, we study the effectiveness of $S_{12}^c$ in distinguishing the $\widetilde{\rm SP}$ scenario from SM and the other NP scenarios. We present this observable separately because it is not directly available in {\tt flavio}.

We calculate $S_{12}^c$ following the expressions in \cite{Altmannshofer:2008dz} and taking the $B\to K^*$ form factors from \cite{Bharucha:2015bzk}. 
We consider the scenario where all the NP WCs are real. It is observed that $S_{12}^c$ by itself will not be able to distinguish the $\widetilde{\rm SP}$ scenario from SP, though it can distinguish it from VA. The combination of $S_{12}^c$ and $S_{6}^c$ could offer an advantage over these two observables separately in certain regions of parameter space. This may be seen in Fig.~\ref{fig:S12combs}.

It may be worthwhile to comment on the observable $S_{12}^c$ in the context of $b\to s \mu\mu$. In this case, the contributions from $C_{P-}$ and the vector WCs are strongly suppressed, i.e., by $m_\mu^2/q^2\sim 1/1500$. However, since the NP WCs in $b\to s \mu\mu$ are severely constrained~\cite{Beaujean:2015gba}, it would not be possible to identify the beyond-SMEFT scenario in the muon mode with the analysis proposed in this article.

\providecommand{\href}[2]{#2}\begingroup\raggedright\endgroup


\begin{thebibliography}{10}
	
	\bibitem{Grzadkowski:2010es}
	B.~Grzadkowski, M.~Iskrzynski, M.~Misiak and J.~Rosiek, \emph{{Dimension-Six
	Terms in the Standard Model Lagrangian}},
	\href{https://doi.org/10.1007/JHEP10(2010)085}{\emph{JHEP} {\bfseries 10}
	(2010) 085} [\href{https://arxiv.org/abs/1008.4884}{{\tt arXiv:1008.4884}}].
	
	\bibitem{Isidori:2023pyp}
	G.~Isidori, F.~Wilsch and D.~Wyler, \emph{{The Standard Model effective field
	theory at work}},  [\href{https://arxiv.org/abs/2303.16922}{{\tt
	arXiv:2303.16922}}].
	
	\bibitem{Buchalla:1995vs}
	G.~Buchalla, A.J.~Buras and M.E.~Lautenbacher, \emph{{Weak decays beyond
	leading logarithms}},
	\href{https://doi.org/10.1103/RevModPhys.68.1125}{\emph{Rev. Mod. Phys.}
	{\bfseries 68} (1996) 1125} [\href{https://arxiv.org/abs/hep-ph/9512380}{{\tt
	hep-ph/9512380}}].
	\bibitem{Jenkins:2017jig}
	E.E.~Jenkins, A.V.~Manohar and P.~Stoffer, \emph{{Low-Energy Effective Field
	Theory below the Electroweak Scale: Operators and Matching}},
	\href{https://doi.org/10.1007/JHEP03(2018)016}{\emph{JHEP} {\bfseries 03}
	(2018) 016} [\href{https://arxiv.org/abs/1709.04486}{{\tt
	arXiv:1709.04486}}].
	\bibitem{Alonso:2014csa}
	R.~Alonso, B.~Grinstein and J.~Martin~Camalich, \emph{{$SU(2)\times U(1)$ gauge
	invariance and the shape of new physics in rare $B$ decays}},
	\href{https://doi.org/10.1103/PhysRevLett.113.241802}{\emph{Phys. Rev. Lett.}
	{\bfseries 113} (2014) 241802} [\href{https://arxiv.org/abs/1407.7044}{{\tt
	arXiv:1407.7044}}].
	
	\bibitem{Cata:2015lta}
	O.~Cat\`a and M.~Jung, \emph{{Signatures of a nonstandard Higgs boson from
	flavor physics}},
	\href{https://doi.org/10.1103/PhysRevD.92.055018}{\emph{Phys. Rev. D}
	{\bfseries 92} (2015) 055018} [\href{https://arxiv.org/abs/1505.05804}{{\tt
	arXiv:1505.05804}}].
	
	\bibitem{Azatov:2018knx}
	A.~Azatov, D.~Bardhan, D.~Ghosh, F.~Sgarlata and E.~Venturini, \emph{{Anatomy
	of $b \to c \tau \nu$ anomalies}},
	\href{https://doi.org/10.1007/JHEP11(2018)187}{\emph{JHEP} {\bfseries 11}
	(2018) 187} [\href{https://arxiv.org/abs/1805.03209}{{\tt
	arXiv:1805.03209}}].
	
	\bibitem{Fuentes-Martin:2020lea}
	J.~Fuentes-Martin, A.~Greljo, J.~Martin~Camalich and J.D.~Ruiz-Alvarez,
	\emph{{Charm physics confronts high-p$_{T}$ lepton tails}},
	\href{https://doi.org/10.1007/JHEP11(2020)080}{\emph{JHEP} {\bfseries 11}
	(2020) 080} [\href{https://arxiv.org/abs/2003.12421}{{\tt
	arXiv:2003.12421}}].
	
	\bibitem{Bause:2020auq}
	R.~Bause, H.~Gisbert, M.~Golz and G.~Hiller, \emph{{Lepton universality and
	lepton flavor conservation tests with dineutrino modes}},
	\href{https://doi.org/10.1140/epjc/s10052-022-10113-6}{\emph{Eur. Phys. J. C}
	{\bfseries 82} (2022) 164} [\href{https://arxiv.org/abs/2007.05001}{{\tt
	arXiv:2007.05001}}].
	
	\bibitem{Bause:2020xzj}
	R.~Bause, H.~Gisbert, M.~Golz and G.~Hiller, \emph{{Rare charm ${c\to
		u\,\nu\bar{\nu}}$ dineutrino null tests for $e^+e^-$ machines}},
	\href{https://doi.org/10.1103/PhysRevD.103.015033}{\emph{Phys. Rev. D}
	{\bfseries 103} (2021) 015033} [\href{https://arxiv.org/abs/2010.02225}{{\tt
	arXiv:2010.02225}}].
	
	\bibitem{Bissmann:2020mfi}
	S.~Bi\ss{}mann, C.~Grunwald, G.~Hiller and K.~Kr\"oninger, \emph{{Top and
	Beauty synergies in SMEFT-fits at present and future colliders}},
	\href{https://doi.org/10.1007/JHEP06(2021)010}{\emph{JHEP} {\bfseries 06}
	(2021) 010} [\href{https://arxiv.org/abs/2012.10456}{{\tt
	arXiv:2012.10456}}].
	
	\bibitem{Bause:2021cna}
	R.~Bause, H.~Gisbert, M.~Golz and G.~Hiller, \emph{{Interplay of dineutrino
	modes with semileptonic rare B-decays}},
	\href{https://doi.org/10.1007/JHEP12(2021)061}{\emph{JHEP} {\bfseries 12}
	(2021) 061} [\href{https://arxiv.org/abs/2109.01675}{{\tt
	arXiv:2109.01675}}].
	
	\bibitem{Bause:2021ihn}
	R.~Bause, H.~Gisbert-Mullor, M.~Golz and G.~Hiller, \emph{{Dineutrino modes
	probing lepton flavor violation}},
	\href{https://doi.org/10.22323/1.398.0563}{\emph{PoS} {\bfseries EPS-HEP2021}
	(2022) 563} [\href{https://arxiv.org/abs/2110.08795}{{\tt
	arXiv:2110.08795}}].
	
	\bibitem{Bruggisser:2021duo}
	S.~Bruggisser, R.~Sch\"afer, D.~van Dyk and S.~Westhoff, \emph{{The Flavor of
	UV Physics}}, \href{https://doi.org/10.1007/JHEP05(2021)257}{\emph{JHEP}
	{\bfseries 05} (2021) 257} [\href{https://arxiv.org/abs/2101.07273}{{\tt
	arXiv:2101.07273}}].
	
	\bibitem{Bause:2022rrs}
	R.~Bause, H.~Gisbert, M.~Golz and G.~Hiller, \emph{{Model-independent analysis
	of $b \rightarrow d$ processes}},
	\href{https://doi.org/10.1140/epjc/s10052-023-11586-9}{\emph{Eur. Phys. J. C}
	{\bfseries 83} (2023) 419} [\href{https://arxiv.org/abs/2209.04457}{{\tt
	arXiv:2209.04457}}].
	
	\bibitem{Sun:2023cuf}
	S.~Sun, Q.-S.~Yan, X.~Zhao and Z.~Zhao, \emph{{Constraining rare B decays by
	\ensuremath{\mu}+\ensuremath{\mu}-\textrightarrow{}tc at future lepton
	colliders}}, \href{https://doi.org/10.1103/PhysRevD.108.075016}{\emph{Phys.
	Rev. D} {\bfseries 108} (2023) 075016}
	[\href{https://arxiv.org/abs/2302.01143}{{\tt arXiv:2302.01143}}].
	
	\bibitem{Grunwald:2023nli}
	C.~Grunwald, G.~Hiller, K.~Kr\"oninger and L.~Nollen, \emph{{More synergies
	from beauty, top, Z and Drell-Yan measurements in SMEFT}},
	\href{https://doi.org/10.1007/JHEP11(2023)110}{\emph{JHEP} {\bfseries 11}
	(2023) 110} [\href{https://arxiv.org/abs/2304.12837}{{\tt
	arXiv:2304.12837}}].
	
	\bibitem{Greljo:2023bab}
	A.~Greljo, J.~Salko, A.~Smolkovi\v{c} and P.~Stangl, \emph{{SMEFT restrictions
	on exclusive b \textrightarrow{} u\ensuremath{\ell}\ensuremath{\nu} decays}},
	\href{https://doi.org/10.1007/JHEP11(2023)023}{\emph{JHEP} {\bfseries 11}
	(2023) 023} [\href{https://arxiv.org/abs/2306.09401}{{\tt
	arXiv:2306.09401}}].
	
	\bibitem{Fajfer:2012vx}
	S.~Fajfer, J.F.~Kamenik and I.~Nisandzic, \emph{{On the $B \to D^* \tau \bar
	\nu_{\tau}$ Sensitivity to New Physics}},
	\href{https://doi.org/10.1103/PhysRevD.85.094025}{\emph{Phys. Rev. D}
	{\bfseries 85} (2012) 094025} [\href{https://arxiv.org/abs/1203.2654}{{\tt
	arXiv:1203.2654}}].
	
	\bibitem{Bause:2023mfe}
	R.~Bause, H.~Gisbert and G.~Hiller, \emph{{Implications of an enhanced
	B\textrightarrow{}K\ensuremath{\nu}\ensuremath{\nu}\textasciimacron{}
	branching ratio}},
	\href{https://doi.org/10.1103/PhysRevD.109.015006}{\emph{Phys. Rev. D}
	{\bfseries 109} (2024) 015006} [\href{https://arxiv.org/abs/2309.00075}{{\tt
	arXiv:2309.00075}}].
	
	\bibitem{Bhattacharya:2023beo}
	S.~Bhattacharya, S.~Jahedi, S.~Nandi and A.~Sarkar, \emph{{Probing flavour
	constrained SMEFT operators through $tc$ production at the Muon collider}},
	[\href{https://arxiv.org/abs/2312.14872}{{\tt arXiv:2312.14872}}].
	
	\bibitem{Chen:2024jlj}
	F.-Z.~Chen, Q.~Wen and F.~Xu, \emph{{Correlating $B\to K^{(\ast)} \nu\bar{\nu}$
	and flavor anomalies in SMEFT}},
	[\href{https://arxiv.org/abs/2401.11552}{{\tt arXiv:2401.11552}}].
	
	\bibitem{Fernandez-Martinez:2024bxg}
	E.~Fern\'andez-Mart\'\i{}nez, X.~Marcano and D.~Naredo-Tuero, \emph{{Global
	Lepton Flavour Violating Constraints on New Physics}},
	[\href{https://arxiv.org/abs/2403.09772}{{\tt arXiv:2403.09772}}].
	
	\bibitem{Karmakar:2024gla}
	S.~Karmakar, A.~Dighe and R.S.~Gupta, \emph{{SMEFT predictions for semileptonic
	processes}},  [\href{https://arxiv.org/abs/2404.10061}{{\tt
	arXiv:2404.10061}}].
	
	\bibitem{Alonso:2012px}
	R.~Alonso, M.B.~Gavela, L.~Merlo, S.~Rigolin and J.~Yepes, \emph{{The Effective
	Chiral Lagrangian for a Light Dynamical ''Higgs Particle''}},
	\href{https://doi.org/10.1016/j.physletb.2013.04.037}{\emph{Phys. Lett. B}
	{\bfseries 722} (2013) 330} [\href{https://arxiv.org/abs/1212.3305}{{\tt
	arXiv:1212.3305}}].
	
	\bibitem{Buchalla:2013rka}
	G.~Buchalla, O.~Cat\`a and C.~Krause, \emph{{Complete Electroweak Chiral
	Lagrangian with a Light Higgs at NLO}},
	\href{https://doi.org/10.1016/j.nuclphysb.2014.01.018}{\emph{Nucl. Phys. B}
	{\bfseries 880} (2014) 552} [\href{https://arxiv.org/abs/1307.5017}{{\tt
	arXiv:1307.5017}}].
	
	\bibitem{Pich:2016lew}
	A.~Pich, I.~Rosell, J.~Santos and J.J.~Sanz-Cillero, \emph{{Fingerprints of
	heavy scales in electroweak effective Lagrangians}},
	\href{https://doi.org/10.1007/JHEP04(2017)012}{\emph{JHEP} {\bfseries 04}
	(2017) 012} [\href{https://arxiv.org/abs/1609.06659}{{\tt
	arXiv:1609.06659}}].
	
	\bibitem{Cohen:2020xca}
	T.~Cohen, N.~Craig, X.~Lu and D.~Sutherland, \emph{{Is SMEFT Enough?}},
	\href{https://doi.org/10.1007/JHEP03(2021)237}{\emph{JHEP} {\bfseries 03}
	(2021) 237} [\href{https://arxiv.org/abs/2008.08597}{{\tt
	arXiv:2008.08597}}].
	
	\bibitem{Burgess:2021ylu}
	C.P.~Burgess, S.~Hamoudou, J.~Kumar and D.~London, \emph{{Beyond the standard
	model effective field theory with $B \rightarrow c \tau^- \overline{\nu}$}},
	\href{https://doi.org/10.1103/PhysRevD.105.073008}{\emph{Phys. Rev. D}
	{\bfseries 105} (2022) 073008} [\href{https://arxiv.org/abs/2111.07421}{{\tt
	arXiv:2111.07421}}].
	
	\bibitem{DiMicco:2019ngk}
	J.~Alison et~al., \emph{{Higgs boson potential at colliders: Status and
	perspectives}}, \href{https://doi.org/10.1016/j.revip.2020.100045}{\emph{Rev.
	Phys.} {\bfseries 5} (2020) 100045}
	[\href{https://arxiv.org/abs/1910.00012}{{\tt arXiv:1910.00012}}].
	
	\bibitem{BaBar:2012obs}
	{\scshape BaBar} collaboration, \emph{{Evidence for an excess of $\bar{B} \to
	D^{(*)} \tau^-\bar{\nu}_\tau$ decays}},
	\href{https://doi.org/10.1103/PhysRevLett.109.101802}{\emph{Phys. Rev. Lett.}
	{\bfseries 109} (2012) 101802} [\href{https://arxiv.org/abs/1205.5442}{{\tt
	arXiv:1205.5442}}].
	
	\bibitem{BaBar:2013mob}
	{\scshape BaBar} collaboration, \emph{{Measurement of an Excess of $\bar{B} \to
	D^{(*)}\tau^- \bar{\nu}_\tau$ Decays and Implications for Charged Higgs
	Bosons}}, \href{https://doi.org/10.1103/PhysRevD.88.072012}{\emph{Phys. Rev.
	D} {\bfseries 88} (2013) 072012} [\href{https://arxiv.org/abs/1303.0571}{{\tt
	arXiv:1303.0571}}].
	
	\bibitem{Belle:2015qfa}
	{\scshape Belle} collaboration, \emph{{Measurement of the branching ratio of
	$\bar{B} \to D^{(\ast)} \tau^- \bar{\nu}_\tau$ relative to $\bar{B} \to
	D^{(\ast)} \ell^- \bar{\nu}_\ell$ decays with hadronic tagging at Belle}},
	\href{https://doi.org/10.1103/PhysRevD.92.072014}{\emph{Phys. Rev. D}
	{\bfseries 92} (2015) 072014} [\href{https://arxiv.org/abs/1507.03233}{{\tt
	arXiv:1507.03233}}].
	
	\bibitem{LHCb:2015gmp}
	{\scshape LHCb} collaboration, \emph{{Measurement of the ratio of branching
	fractions $\mathcal{B}(\bar{B}^0 \to
	D^{*+}\tau^{-}\bar{\nu}_{\tau})/\mathcal{B}(\bar{B}^0 \to
	D^{*+}\mu^{-}\bar{\nu}_{\mu})$}},
	\href{https://doi.org/10.1103/PhysRevLett.115.111803}{\emph{Phys. Rev. Lett.}
	{\bfseries 115} (2015) 111803} [\href{https://arxiv.org/abs/1506.08614}{{\tt
	arXiv:1506.08614}}].
	
	\bibitem{LHCb:2017vlu}
	{\scshape LHCb} collaboration, \emph{{Measurement of the ratio of branching
	fractions
	$\mathcal{B}(B_c^+\,\to\,J/\psi\tau^+\nu_\tau)$/$\mathcal{B}(B_c^+\,\to\,J/\psi\mu^+\nu_\mu)$}},
	\href{https://doi.org/10.1103/PhysRevLett.120.121801}{\emph{Phys. Rev. Lett.}
	{\bfseries 120} (2018) 121801} [\href{https://arxiv.org/abs/1711.05623}{{\tt
	arXiv:1711.05623}}].
	
	\bibitem{LHCb:2014cxe}
	{\scshape LHCb} collaboration, \emph{{Differential branching fractions and
	isospin asymmetries of $B \to K^{(*)} \mu^+ \mu^-$ decays}},
	\href{https://doi.org/10.1007/JHEP06(2014)133}{\emph{JHEP} {\bfseries 06}
	(2014) 133} [\href{https://arxiv.org/abs/1403.8044}{{\tt arXiv:1403.8044}}].
	
	\bibitem{LHCb:2022zom}
	{\scshape LHCb} collaboration, \emph{{Measurement of lepton universality
	parameters in $B^+\to K^+\ell^+\ell^-$ and $B^0\to K^{*0}{\ell}^+{\ell}^-$
	decays}},  [\href{https://arxiv.org/abs/2212.09153}{{\tt arXiv:2212.09153}}].
	
	\bibitem{Belle-II:2023esi}
	{\scshape Belle-II} collaboration, \emph{{Evidence for $B^{+}\to
	K^{+}\nu\bar{\nu}$ Decays}},  [\href{https://arxiv.org/abs/2311.14647}{{\tt
	arXiv:2311.14647}}].
	
	\bibitem{LHCb:2013ghj}
	{\scshape LHCb} collaboration, \emph{{Measurement of Form-Factor-Independent
	Observables in the Decay $B^{0} \to K^{*0} \mu^+ \mu^-$}},
	\href{https://doi.org/10.1103/PhysRevLett.111.191801}{\emph{Phys. Rev. Lett.}
	{\bfseries 111} (2013) 191801} [\href{https://arxiv.org/abs/1308.1707}{{\tt
	arXiv:1308.1707}}].
	
	\bibitem{Descotes-Genon:2012isb}
	S.~Descotes-Genon, J.~Matias, M.~Ramon and J.~Virto, \emph{{Implications from
	clean observables for the binned analysis of $B -> K*\mu^+\mu^-$ at large
	recoil}}, \href{https://doi.org/10.1007/JHEP01(2013)048}{\emph{JHEP}
	{\bfseries 01} (2013) 048} [\href{https://arxiv.org/abs/1207.2753}{{\tt
	arXiv:1207.2753}}].
	
	\bibitem{Descotes-Genon:2013wba}
	S.~Descotes-Genon, J.~Matias and J.~Virto, \emph{{Understanding the $B\to
	K^*\mu^+\mu^-$ Anomaly}},
	\href{https://doi.org/10.1103/PhysRevD.88.074002}{\emph{Phys. Rev. D}
	{\bfseries 88} (2013) 074002} [\href{https://arxiv.org/abs/1307.5683}{{\tt
	arXiv:1307.5683}}].
	
	\bibitem{Karmakar:2023rdt}
	S.~Karmakar, S.~Chattopadhyay and A.~Dighe, \emph{{Identifying physics beyond
	SMEFT in the angular distribution of
	\ensuremath{\Lambda}b\textrightarrow{}\ensuremath{\Lambda}c(\textrightarrow{}\ensuremath{\Lambda}\ensuremath{\pi})\ensuremath{\tau}\ensuremath{\nu}\textasciimacron{}\ensuremath{\tau}
	decay}}, \href{https://doi.org/10.1103/PhysRevD.110.015010}{\emph{Phys. Rev.
	D} {\bfseries 110} (2024) 015010}
	[\href{https://arxiv.org/abs/2305.16007}{{\tt arXiv:2305.16007}}].
	
	\bibitem{Alonso:2015sja}
	R.~Alonso, B.~Grinstein and J.~Martin~Camalich, \emph{{Lepton universality
	violation and lepton flavor conservation in $B$-meson decays}},
	\href{https://doi.org/10.1007/JHEP10(2015)184}{\emph{JHEP} {\bfseries 10}
	(2015) 184} [\href{https://arxiv.org/abs/1505.05164}{{\tt
	arXiv:1505.05164}}].
	
	\bibitem{Crivellin:2017zlb}
	A.~Crivellin, D.~M\"uller and T.~Ota, \emph{{Simultaneous explanation of
	R(D$^{(*)}$) and b\textrightarrow{}s\ensuremath{\mu}$^{+}$
	\ensuremath{\mu}$^{-}$: the last scalar leptoquarks standing}},
	\href{https://doi.org/10.1007/JHEP09(2017)040}{\emph{JHEP} {\bfseries 09}
	(2017) 040} [\href{https://arxiv.org/abs/1703.09226}{{\tt
	arXiv:1703.09226}}].
	
	\bibitem{Calibbi:2017qbu}
	L.~Calibbi, A.~Crivellin and T.~Li, \emph{{Model of vector leptoquarks in view
	of the $B$-physics anomalies}},
	\href{https://doi.org/10.1103/PhysRevD.98.115002}{\emph{Phys. Rev. D}
	{\bfseries 98} (2018) 115002} [\href{https://arxiv.org/abs/1709.00692}{{\tt
	arXiv:1709.00692}}].
	
	\bibitem{Capdevila:2017iqn}
	B.~Capdevila, A.~Crivellin, S.~Descotes-Genon, L.~Hofer and J.~Matias,
	\emph{{Searching for New Physics with $b\to s\tau^+\tau^-$ processes}},
	\href{https://doi.org/10.1103/PhysRevLett.120.181802}{\emph{Phys. Rev. Lett.}
	{\bfseries 120} (2018) 181802} [\href{https://arxiv.org/abs/1712.01919}{{\tt
	arXiv:1712.01919}}].
	
	\bibitem{Allwicher:2023xba}
	L.~Allwicher, D.~Becirevic, G.~Piazza, S.~Rosauro-Alcaraz and O.~Sumensari,
	\emph{{Understanding the first measurement of
	B(B\textrightarrow{}K\ensuremath{\nu}\ensuremath{\nu}\textasciimacron{})}},
	\href{https://doi.org/10.1016/j.physletb.2023.138411}{\emph{Phys. Lett. B}
	{\bfseries 848} (2024) 138411} [\href{https://arxiv.org/abs/2309.02246}{{\tt
	arXiv:2309.02246}}].
	
	\bibitem{LHCb:2017myy}
	{\scshape LHCb} collaboration, \emph{{Search for the decays
	$B_s^0\to\tau^+\tau^-$ and $B^0\to\tau^+\tau^-$}},
	\href{https://doi.org/10.1103/PhysRevLett.118.251802}{\emph{Phys. Rev. Lett.}
	{\bfseries 118} (2017) 251802} [\href{https://arxiv.org/abs/1703.02508}{{\tt
	arXiv:1703.02508}}].
	
	\bibitem{BaBar:2016wgb}
	{\scshape BaBar} collaboration, \emph{{Search for $B^{+}\rightarrow K^{+}
	\tau^{+}\tau^{-}$ at the BaBar experiment}},
	\href{https://doi.org/10.1103/PhysRevLett.118.031802}{\emph{Phys. Rev. Lett.}
	{\bfseries 118} (2017) 031802} [\href{https://arxiv.org/abs/1605.09637}{{\tt
	arXiv:1605.09637}}].
	
	\bibitem{Belle:2021ecr}
	{\scshape Belle} collaboration, \emph{{Search for the decay
	B0\textrightarrow{}K*0\ensuremath{\tau}+\ensuremath{\tau}- at the Belle
	experiment}}, \href{https://doi.org/10.1103/PhysRevD.108.L011102}{\emph{Phys.
	Rev. D} {\bfseries 108} (2023) L011102}
	[\href{https://arxiv.org/abs/2110.03871}{{\tt arXiv:2110.03871}}].
	
	\bibitem{Straub:2018kue}
	D.M.~Straub, \emph{{flavio: a Python package for flavour and precision
	phenomenology in the Standard Model and beyond}},
	[\href{https://arxiv.org/abs/1810.08132}{{\tt arXiv:1810.08132}}].
	
	\bibitem{Belle-II:2018jsg}
	{\scshape Belle-II} collaboration, \emph{{The Belle II Physics Book}},
	\href{https://doi.org/10.1093/ptep/ptz106}{\emph{PTEP} {\bfseries 2019}
	(2019) 123C01} [\href{https://arxiv.org/abs/1808.10567}{{\tt
	arXiv:1808.10567}}].
	
	\bibitem{LHCb:2018roe}
	{\scshape LHCb} collaboration, \emph{{Physics case for an LHCb Upgrade II -
	Opportunities in flavour physics, and beyond, in the HL-LHC era}},
	[\href{https://arxiv.org/abs/1808.08865}{{\tt arXiv:1808.08865}}].
	
	\bibitem{FCC:2018byv}
	{\scshape FCC} collaboration, \emph{{FCC Physics Opportunities}: {Future
	Circular Collider Conceptual Design Report Volume 1}},
	\href{https://doi.org/10.1140/epjc/s10052-019-6904-3}{\emph{Eur. Phys. J. C}
	{\bfseries 79} (2019) 474}.
	
	\bibitem{Apollonio:2019zjt}
	A.~Apollonio et~al., \emph{{FCC-ee Operation Model, Availability \&
	Performance}},  in \emph{{62nd ICFA Advanced Beam Dynamics Workshop on High
	Luminosity Circular $e^+ e^-$ Colliders}}, p.~WEPAB03, JACOW, 2019,
	\href{https://doi.org/10.18429/JACoW-eeFACT2018-WEPAB03}{DOI}.
	
	\bibitem{Aliev:2000ae}
	T.M.~Aliev and M.~Savci, \emph{{Lepton polarization and CP violating effects in
	$B\to K^*\tau^+\tau^-$ decay in standard and two Higgs doublet
	models}}, \href{https://doi.org/10.1016/S0370-2693(00)00454-8}{\emph{Phys.
	Lett. B} {\bfseries 481} (2000) 275}
	[\href{https://arxiv.org/abs/hep-ph/0003188}{{\tt hep-ph/0003188}}].
	
	\bibitem{Choudhury:2003mi}
	S.R.~Choudhury, N.~Gaur, A.S.~Cornell and G.C.~Joshi, \emph{{Lepton
	polarization correlations in $B\to K^*\tau^+\tau^-$}},
	\href{https://doi.org/10.1103/PhysRevD.68.054016}{\emph{Phys. Rev. D}
	{\bfseries 68} (2003) 054016}
	[\href{https://arxiv.org/abs/hep-ph/0304084}{{\tt hep-ph/0304084}}].
	
	\bibitem{SinghChundawat:2022ldm}
	N.R.~Singh~Chundawat, \emph{{New physics in
	$B\to K^*\tau^+\tau^-$: A model independent
	analysis}}, \href{https://doi.org/10.1103/PhysRevD.107.055004}{\emph{Phys.
	Rev. D} {\bfseries 107} (2023) 055004}
	[\href{https://arxiv.org/abs/2212.01229}{{\tt arXiv:2212.01229}}].
	
	\bibitem{Das:2018iap}
	D.~Das, \emph{{On the angular distribution of $\Lambda_b\to\Lambda(\to
	N\pi)\tau^+\tau^-$ decay}},
	\href{https://doi.org/10.1007/JHEP07(2018)063}{\emph{JHEP} {\bfseries 07}
	(2018) 063} [\href{https://arxiv.org/abs/1804.08527}{{\tt
	arXiv:1804.08527}}].
	
	\bibitem{Aebischer:2015fzz}
	J.~Aebischer, A.~Crivellin, M.~Fael and C.~Greub, \emph{{Matching of gauge
	invariant dimension-six operators for $b\to s$ and $b\to c$ transitions}},
	\href{https://doi.org/10.1007/JHEP05(2016)037}{\emph{JHEP} {\bfseries 05}
	(2016) 037} [\href{https://arxiv.org/abs/1512.02830}{{\tt
	arXiv:1512.02830}}].
	
	\bibitem{Bobeth:2011st}
	C.~Bobeth and U.~Haisch, \emph{{New Physics in $\Gamma_{12}^s$: ($\bar{s}
	b$)$(\bar{\tau} \tau)$ Operators}},
	\href{https://doi.org/10.5506/APhysPolB.44.127}{\emph{Acta Phys. Polon. B}
	{\bfseries 44} (2013) 127} [\href{https://arxiv.org/abs/1109.1826}{{\tt
	arXiv:1109.1826}}].
	
	\bibitem{Altmannshofer:2008dz}
	W.~Altmannshofer, P.~Ball, A.~Bharucha, A.J.~Buras, D.M.~Straub and M.~Wick,
	\emph{{Symmetries and Asymmetries of $B \to K^{*} \mu^{+} \mu^{-}$ Decays in
	the Standard Model and Beyond}},
	\href{https://doi.org/10.1088/1126-6708/2009/01/019}{\emph{JHEP} {\bfseries
	01} (2009) 019} [\href{https://arxiv.org/abs/0811.1214}{{\tt
	arXiv:0811.1214}}].
	
	\bibitem{Bharucha:2015bzk}
	A.~Bharucha, D.M.~Straub and R.~Zwicky, \emph{{$B\to V\ell^+\ell^-$ in the
	Standard Model from light-cone sum rules}},
	\href{https://doi.org/10.1007/JHEP08(2016)098}{\emph{JHEP} {\bfseries 08}
	(2016) 098} [\href{https://arxiv.org/abs/1503.05534}{{\tt
	arXiv:1503.05534}}].
	
	\bibitem{Beaujean:2015gba}
	F.~Beaujean, C.~Bobeth and S.~Jahn, \emph{{Constraints on tensor and scalar
	couplings from $B\rightarrow K\bar{\mu }\mu $ and $B_s\rightarrow \bar{\mu
	}\mu $}}, \href{https://doi.org/10.1140/epjc/s10052-015-3676-2}{\emph{Eur.
	Phys. J. C} {\bfseries 75} (2015) 456}
	[\href{https://arxiv.org/abs/1508.01526}{{\tt arXiv:1508.01526}}].
	
\end{thebibliography}
\end{document}